\documentclass[preprint,aps]{revtex4}
\usepackage[latin9]{luainputenc}
\usepackage{float}
\usepackage{amsmath}
\usepackage{color}
\usepackage{amssymb}
\usepackage{graphicx}
\usepackage{setspace}

\usepackage{media9} 
\usepackage{hyperref}

\makeatletter

\@ifundefined{textcolor}{}
{%
 \definecolor{BLACK}{gray}{0}
 \definecolor{WHITE}{gray}{1}
 \definecolor{RED}{rgb}{1,0,0}
 \definecolor{GREEN}{rgb}{0,1,0}
 \definecolor{BLUE}{rgb}{0,0,1}
 \definecolor{CYAN}{cmyk}{1,0,0,0}
 \definecolor{MAGENTA}{cmyk}{0,1,0,0}
 \definecolor{YELLOW}{cmyk}{0,0,1,0}
}

\usepackage{amsfonts}
\usepackage{bm}
\usepackage[dvips]{epsfig}
\usepackage[dvips]{graphicx}
\usepackage{float}
\usepackage{dcolumn}
\usepackage{subfigure}
\usepackage{ifpdf}
\usepackage{multirow}

\newcommand{\be}{\begin{equation}}
\newcommand{\ee}{\end{equation}}
\newcommand{\bes}{\begin{subequations}}
\newcommand{\ees}{\end{subequations}}
\newcommand{\ben}{\begin{eqnarray}}
\newcommand{\een}{\end{eqnarray}}

\makeatother

\begin{document}

\title{{Scattering of sine-Gordon kinks with internal structure in an extended nonlinear $O(3)$ sigma model}}
\author{ Adalto R. Gomes and Fabiano C. Simas}
\email{adalto.gomes@ufma.br, fc.simas@ufma.br}


\affiliation{
Programa de P\'os-Gradua\c c\~ao em F\'\i sica, Universidade Federal do Maranh\~ao\\Campus Universit\'ario do Bacanga, 65085-580, S\~ao Lu\'\i s, Maranh\~ao, Brazil\\
Departamento de F\'isica, Universidade Federal do Maranh\~ao (UFMA), Campus Universit\'ario do Bacanga, 65085-580, S\~ao Lu\'is, Maranh\~ao, Brazil
}

\begin{abstract}
In this paper, we present topological defects in $(1,1)$ dimensions, described by an extended nonlinear $O(3)$ sigma model. We consider spherical coordinates $(\phi,\chi)$ in the $S^2$ isotopic space and a potential $V(\phi,\chi)$. For specific forms of the potential, it is possible to apply the Bogomol'nyi method, resulting in first-order equations of motion with solutions that minimize energy. We study a model with an explicit solution for the field $\phi$ that resembles the ubiquitous sine-Gordon kink/antikink, but with an internal structure given by the field $\chi$ with a form antilump/lump that depends on a constant $C$. The soliton-antisoliton scattering process depends on $C$ and on the initial velocity of the pair. 
Some results are reported, such as:
one-bounce scattering for $\phi$, or strong emission of radiation for $\phi$, followed by  i) annihilation of $\chi$; ii) same pattern antilump-antilump or lump-antilump for $\chi$; iii) inversion antilump-antilump to lump-lump for $\chi$; iv) inversion antilump-lump to lump-antilump for $\chi$.
Other findings are: v) annihilation of the pair soliton-antisoliton with the emission of scalar radiation; vi) emission of pairs of oscillations around the vacuum for $\phi$ and $\chi$. The energy density shows that the defect has an internal structure as a nested defect of an antilump inside a kink. The lump core of the defect is responsible for the emission of radiation. The changing of structure of the defects during the scattering is analyzed not only with the field profiles in the physical $(1,1)$ space, but also in the internal $S^2$ space, which gives some new insight into the process.

\end{abstract}

\maketitle

\section{Introduction}\label{introduction}

Topological defects are field theory solutions with localized density energy that propagate freely without losing form. For each solution, there is a topological map between the physical coordinate space and the internal field space, or space of configurations \cite{mt}. The degree of mapping characterizes the topological charge, associated to a current not related to the Noether theorem. In $(1,1)$ dimensions, we have the kink and antikink as the simplest topological defects. Derrick's \cite{der} theorem, under specific requirements, forbids static solutions constructed only with scalar fields in more than two spatial dimensions. One way to evade this is to consider a multiple of fields with a geometric constraint, as done, for instance, in the nonlinear $O(3)$ sigma model. In this model, the Lagrangian of free fields is supplemented by a geometrical constraint that results, after eliminating one of the field components, in a nontrivial coupling \cite{shif}. Physical problems studied with the nonlinear $O(3)$ sigma model include: spin dynamics in Heisenberg ferromagnets \cite{hald, alon1, alon2} and asymptotic freedom and mass generation \cite{asympt}.  When coupled to gravity, it gives hairy black holes \cite{bh}. The entanglement and R\'enyi entropies of this model were studied in Ref. \cite{ent}.

In $(2,1)$ dimensions, after fixing the vacuum due to spontaneous breaking symmetry, every finite energy field configuration of the $O(3)$ sigma model corresponds to a mapping $S^2\to S^2$ with homotopy group $\pi_2(S^2)=\mathbb{Z}$ between the physical space and the internal field space. The solutions are unstable, due to the conformal invariance of the model in $(2,0)$ dimensions, meaning that the structures can have any size, or the initial size can change due to small perturbations \cite{zak1}. Despite their intrinsic instability, these lump defects do not shrink to the vacuum so fast, and their interaction can be investigated. Indeed, numerical investigation of scattering of these defects, called lumps, was investigated in the Ref. \cite{zak1}, using the Riemann sphere in the configuration space and the complex coordinate in the $xy$ physical plane plus a value $\infty$ for infinity. Then the target sphere manifold $S^2$ turns into a complex projective line, and the model turns to the equivalent $\mathbb{CP}^1$ sigma model. 

The $O(3)$ sigma model in $(1,1)$-dimensions involving an explicitly broken symmetry was investigated in the Ref. \cite{log}. A finite energy configuration corresponds to a mapping $S^1\to S^2$ with a fixed point. It was shown that in the internal space $S^2_{int}$ there are solutions in a form of non-contractible loops beginning and ending at the vacuum point, a necessary condition for sphalerons, unstable saddle point like solutions \cite{log}. The model also has sine-Gordon kink solutions and their non-static generalization \cite{log}. A massive nonlinear $O(3)$ sigma model in $(1,1)$-dimensions was investigated in the Refs. \cite{alon1,alon2}. The quadratic potential was chosen as the simplest one to give mass to the fundamental quanta \cite{alon2}. An interesting aspect of these works is the use of spherical coordinates for the fields, leading to a curved metric for the configuration space. Here the metric components depend on the fields, meaning that there is a geometric constraint. Examples of geometric constraints for kinks can be found in the Refs. \cite{baz1, baz2, baz3, baz4}. The presence of such constrictions leads to a modification of the kinetic term, and consequently to the existence of the internal structure of the profiles. In the Refs. \cite{baz5,marq1}, the authors also discussed geometrically constrained kink configurations with two and three scalar fields, respectively. The Ref. \cite{joao}
  explored how the existence of geometric constrictions affects the scattering process of kink solutions. In the Ref. \cite{jubert}, the formation of a N\'eel-type wall in micrometer-sized Fe${}_{20}$Ni${}_{80}$ elements with a two-kink structure containing geometric constrictions was analyzed using numerical simulations and scanning electron microscopy. Geometric constriction was also considered for bubble universe collisions in the Ref. \cite{green} to verify the free passage hypothesis in the ultra-relativistic limit. In inflationary scenarios on the string landscape, the fields are inserted in manifolds with Calabi-Yau compactifications \cite{infl}. This means that the study of field collisions in a curved space can lead to physically relevant results, such as of bubbles constraining eternal inflation theories \cite{bubb}. Other interesting aspects of the search for new sigma models on the sphere \cite{alon4} are directly related to their possible applications, including spintronics and the process of storing information \cite{chuma,lesne}.   
  Very recently, deformation methods \cite{balomal} have been applied to generalize scalar field theories in Euclidean target spaces to sigma models \cite{alon3}. This allowed deformations of sigma models in the plane $R^2$ and in the sphere $S^2$, as well as the transference of solutions between these two manifolds \cite{alon3}.

In this work we will consider an extension of the $(1,1)$ O(3) sigma model with spontaneous breaking symmetry. This leads to a model with two interacting and geometrically coupled scalar fields. The model is constructed such that one of the fields has the same solution as the $(1,1)$ sine-Gordon model, whereas the other field is responsible for the presence of an internal structure of the defect. The paper is structured as follows: in the next section, we introduce the general model, obtain the equations of motion, and consider stability analysis. In the Sect. III we present our model, obtained after imposing the sine-Gordon solution for the field $\chi$. The solution for the field $\chi$ depends on a parameter that modifies the internal structure. We show the solutions in internal space. In the Sect. IV we present the main scattering results for soliton-antisoliton collisions. In the Sect. V we present the stability analysis for the model. In two important limits, the equations for linear perturbations of the two fields are decoupled, leading to Schrodinger-like equations, which have a more direct interpretation. In the Sect. VI we present our main conclusion. Supplementary material are attached and described in the Sect. VII.    

\section{$S^2$ geometric coupling and a general potential term}
 We start with the $O(3)$ sigma model in $(1,1)$ dimensions, with Minkowski signature $(+,-)$, described by the action \cite{mt}
\be 
S=\int dt dx \biggl[\frac1{2}\partial_\mu\vec \varphi\cdot\partial^\mu\vec\varphi-\nu(1-\vec\varphi\cdot\vec\varphi)\biggr],
\label{act-sigma}
\ee
with $\vec\varphi=(\varphi^1, \varphi^2, \varphi^3)$ and $\nu$ is a Lagrange multiplier. The equations of motion are complex, as a result of the constraint $\vec\varphi\cdot\vec\varphi=1$. Now, let us change variables considering spherical coordinates in the target space of unitary radius:
\begin{eqnarray}
\varphi^1&=&\sin{\chi}\cos{\phi}\\
\varphi^2&=&\sin{\chi}\sin{\phi}\\
\varphi^3&=&\cos{\chi}\\
\end{eqnarray}
 This change the action from Eq. (\ref{act-sigma}) to  \be 
S=\int dt dx \biggl( \frac12\partial_\mu\chi\partial^\mu\chi+\frac12\sin^2\chi\partial_\mu\phi\partial^\mu\phi  \biggl).
\label{act-sigma2}
\ee
Now we have two coupled real scalar fields $(\phi,\chi)$ in an $S^2$ internal space with metric 
\be
g_{ij}=\begin{pmatrix}
1 & 0\\
0 & \sin^2\chi.
\end{pmatrix}
\ee
From the action given by the Eq. (\ref{act-sigma2}) one can see that the coupling is purely geometric, with no potential term. 

In the present work we extend the action from Eq. (\ref{act-sigma2}) including a potential term:
\be
S=\int dt dx \biggl( \frac12\partial_\mu\chi\partial^\mu\chi+\frac12\sin^2\chi\partial_\mu\phi\partial^\mu\phi-V(\phi,\chi)\biggr).
\ee
This means that, in addition to the geometric coupling, we have a direct coupling given by the potential. This results in the set of equations of motion
\ben
\label{eq1}
\partial_\mu(\sin^2\chi\partial^\mu\phi)+V_\phi(\phi,\chi)&=&0,\\
\label{eq2}
\partial_\mu\partial^\mu\chi-\frac12\sin(2\chi)\partial_\mu\phi\partial^\mu\phi+V_\chi(\phi,\chi)&=&0,
\een
where $V_\phi=\partial V/\partial\phi$ and so on.
Static solutions are solutions of the equations
\ben
\frac{d^2\chi}{dx^2}-\frac12\sin(2\chi)\biggl(\frac{d\phi}{dx} \biggr)^2&=&V_\chi(\phi,\chi), \\
\frac{d}{dx}\biggl(\sin^2\chi\frac{d\phi}{dx}\biggr)&=&V_\phi(\phi,\chi).
\een
The energy density is given by
\be
\rho=\frac12\sin^2\chi\biggl(\frac{d\phi}{dx}\biggr)^2 + \frac12\biggl(\frac{d\chi}{dx}\biggr)^2 + V(\phi,\chi).
\ee

The BPS \cite{bps1,bps2} method can be applied for the potential given by
\be
V(\phi,\chi)=\frac12\frac{W_\phi^2}{\sin^2\chi} + \frac12{W_\chi^2}. 
\ee
In this case the energy density is given by  
\be
\rho=\frac12\sin^2\chi\biggl(\frac{d\phi}{dx}\mp\frac{W_\phi}{\sin^2\chi}\biggr)^2 + \frac12\biggl(\frac{d\chi}{dx}\mp W_\chi\biggr)^2 + W_\phi \frac{d\phi}{dx} + W_\chi \frac{d\chi}{dx}.
\ee
Then, for 
\ben 
\label{eom_1a}
\frac{d\phi}{dx}&=&\pm\frac{W_\phi}{\sin^2\chi} \,\, ,\label{1eqphi}\\
\label{eom_1b}
\frac{d\chi}{dx}&=&\pm W_\chi\,\, , \label{1eqchi}
\een
the energy is minimized, and given by $E=|W(x\to+\infty)-W(x\to-\infty)|$.

Stability analysis around the $BPS$ solution $(\phi(x), \chi(x))$ considers that the scalar fields are given by
\be
\begin{pmatrix}
\tilde\phi(x,t)\\
\tilde\chi(x,t)
\end{pmatrix}
=
\begin{pmatrix}
\phi(x)\\
\chi(x)
\end{pmatrix}
+acos(\omega t)
\begin{pmatrix}
\eta_1(x)\\
\eta_2(x)
\end{pmatrix}.
\ee
This and the equations of motion (\ref{eq1}),(\ref{eq2}) give, to first order in the perturbations $(\eta_1,\eta_2)$,
\be
\label{eq_pert}
\widehat{\mathcal{H}} \begin{pmatrix}
\eta_1\\
\eta_2
\end{pmatrix}
=
 \begin{pmatrix}
\sin(\chi)^2\omega^2\eta_1\\
\omega^2\eta_2
\end{pmatrix},
\ee
where
\be
\label{hamil}
\widehat{\mathcal{H}}=\left( {\begin{array}{cc}
    H_{11} & H_{12} \\
    H_{21} & H_{22} \\
  \end{array} } \right)
\ee
and
\ben
H_{11}&=& -\sin(2\chi)\frac{d\chi}{dx} \frac{d}{dx} - \sin^2(\chi)\frac{d^2}{dx^2} + V_{\phi\phi}\,\, , \\
H_{12}&=& -\sin(2\chi)\frac{d\phi}{dx}\frac{d}{dx} -2\cos(2\chi)\frac{d\phi}{dx}\frac{d\chi}{dx} + V_{\phi\chi}\,\, , \\
H_{21}&=& \sin(2\chi)\frac{d\phi}{dx}\frac{d}{dx} + V_{\chi\phi}\,\, , \\
H_{22}&=& -\frac{d^2}{dx^2} + \cos(2\chi)  \biggl(\frac{d\phi}{dx} \biggr)^2 + V_{\chi\chi}\,\, .
\een
In general, the Eq.(\ref{eq_pert}) is a set of two coupled differential equations with no known analytical solution. Also, a non diagonal matrix given by the Eq. (\ref{hamil}) leads to further difficulties in interpreting the results.  However, for some specific cases, the perturbations are decoupled and turned into a Sturm-Liouville problem, which makes the stability analysis easier. This will be elaborated in the Sect. VII for the model presented in the following section. 

\section{An Extended sine-Gordon Model}

An interesting choice for $W$ starts with
\be
\label{wphi}
W_\phi=2\sin\biggl(\frac{\phi}2\biggr)\sin^2\chi.
\ee
\begin{figure}
	\includegraphics[{angle=0,width=12cm}]{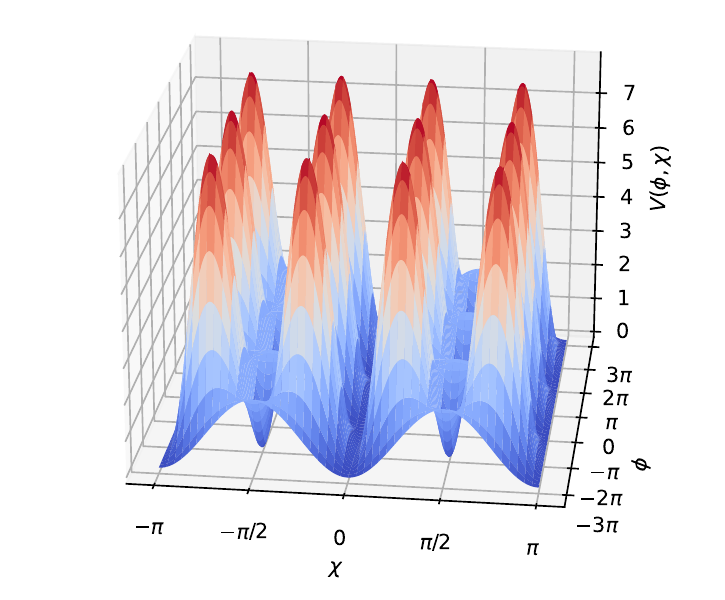}	 
	\caption{Potential $V(\phi,\chi)$ for the extended sine-Gordon model. } 
	\label{fig_V}%
\end{figure}

After integration $W_\phi$ from the Eq. (\ref{wphi}) we attain
\ben
W&=&-4\sin^2(\chi)\cos\biggl(\frac{\phi}2\biggr)+f(\chi),\\
\label{Wchi}
W_\chi&=&-4\sin(2\chi)\cos\biggl(\frac{\phi}2\biggr)+\frac{df}{d\chi}.
\een
For simplicity we chose $f(\chi)=0$. Then we have the following first-order equations for static solutions:
\ben 
\label{eom_1a_model1}
\frac{d\phi}{dx}&=&\pm 2\sin\biggl(\frac{\phi}2\biggr),\\
\label{eom_1b_model1}
\frac{d\chi}{dx}&=&\mp 4\sin(2\chi)\cos\biggl(\frac{\phi}2\biggr).
\een
The potential is given by
\be
V(\phi,\chi)=2 \sin^2\chi  \biggl( \sin^2\biggl(\frac{\phi}2\biggr) + 16 \cos^2\chi \cos^2\biggl(\frac{\phi}2\biggr) \biggr)
\ee
Note that the potential is non-negative. The conditions $V(\phi,\chi)=0$ define the vacuum solutions, given by $\phi=\pm2n\pi$, $\chi=\pm(2n+1)\pi/2$, with $n=0,1,2,...$. In the Fig. \ref{fig_V} one can see six vacuum $\phi=0,\pm 2\pi$ and $\chi=\pm\pi/2$. We are interested in solutions connecting two neighbour vacuum, that is belonging to the same topological sector. The simplest choice to cover the $S^2_{int}$ surface is to restrict the domain of fields to $0\leq\phi\le2\pi$ and $0\leq\chi\leq\pi$.

\begin{figure}
    \includegraphics[{angle=0,width=10cm}]{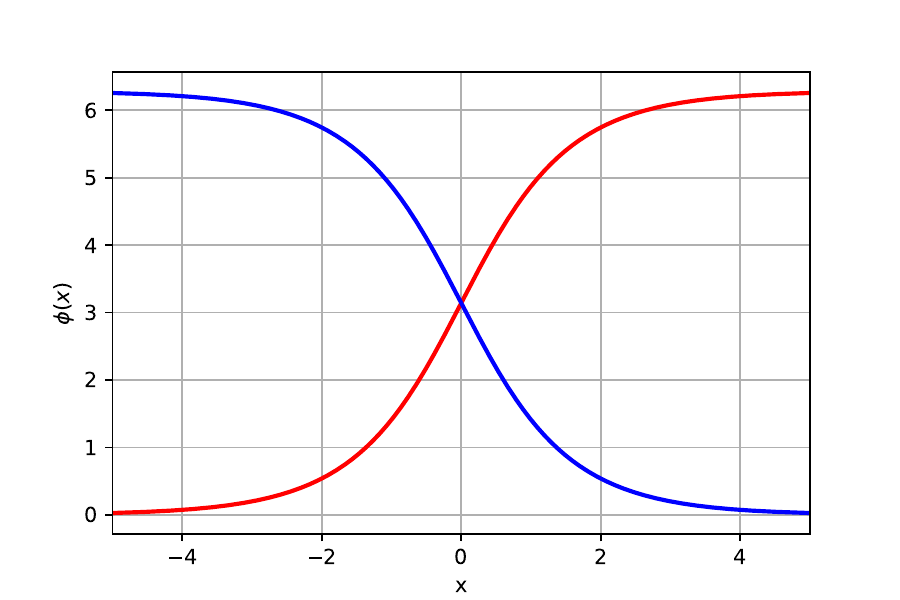}
	\caption{Scalar field $\phi(x)$ for soliton (red) and antisoliton (blue).} 
	\label{fig_phi}%
\end{figure}

We will consider solutions that interpolate between the two vacuum $(\phi,\chi)=(0,\pi/2)\to(2\pi,\pi/2)$ (soliton) and $(\phi,\chi)=(2\pi,\pi/2)\to(0,\pi/2)$ (antisoliton). Let us start from the soliton solution.
From Eq. (\ref{1eqphi}) we have
\be
\label{phix}
\phi(x)= 4\arctan(e^x).
\ee
In the Fig. \ref{fig_phi} we see that the solution for  $\phi(x)$ interpolates between $0$ and $2\pi$, and has the same kink solution for the one-field sine-Gordon model. Now, Eqs. (\ref{eom_1a}) and (\ref{eom_1b}) give
\be
\frac{d\phi}{d\chi}=\frac{W_\phi}{(\sin^2\chi) W_\chi}=-\frac{\tan(\phi/2)}{2\sin(2\chi)}.
\ee
This equation has the solution
\be
\chi(\phi)=\arctan(C\sin^{-8}(\phi/2)),
\ee
with $C$ a real constant. Now, with Eq. ({\ref{phix}}) we get
\be
\chi(C,x)=\arctan(C\cosh^8(x)),
\ee
where we included the explicit dependence in $C$ of the field $\chi(x)$. However, for this solution, the domain of $\chi(x)$ for $C<0$ is out of the desired domain $0\leq\chi\leq\pi$. To fix this we note that the equations of motion are invariant under the transformation $\chi\to\chi+\pi$. Then we can write 
\begin{eqnarray}
    \label{solchi}
\chi(C,x)&=&\arctan(C\cosh^8(x)), \,\,C>0\\
\label{solchi2}
\chi(C,x)&=&\arctan(C\cosh^8(x))+\pi, \,\, C<0.
\end{eqnarray}

\begin{figure}
    \includegraphics[{angle=0,width=8cm, height=6cm}]{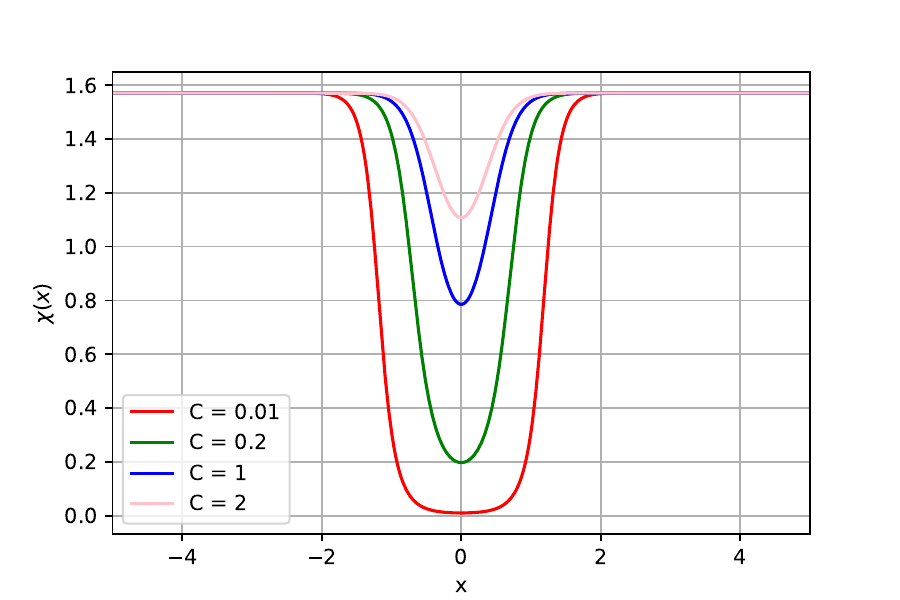}
    \includegraphics[{angle=0,width=8cm, height=6cm}]{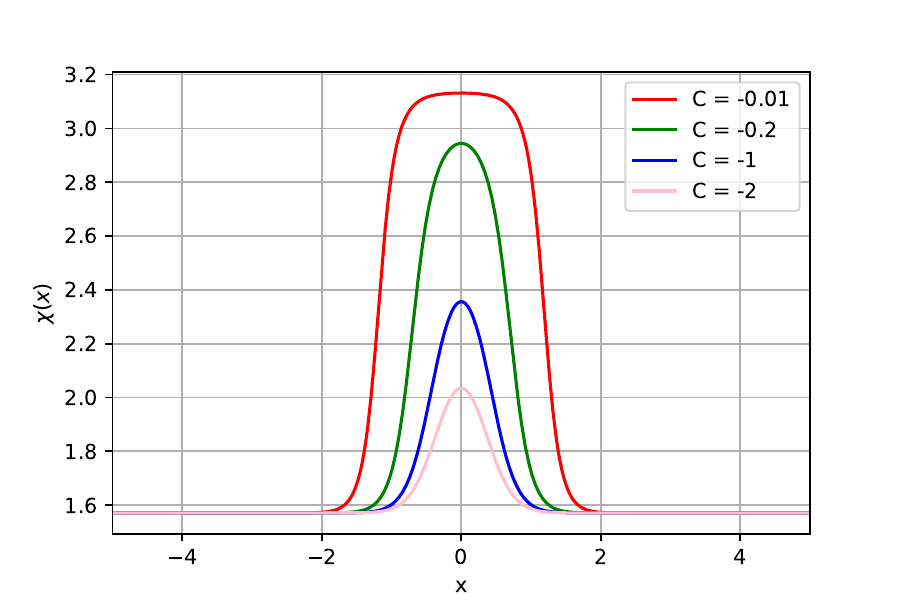}
	\caption{Scalar field $\chi(x)$ for a) $C>0$ and b) $C<0$ ($C=-0.01$ (red), $C=-0.2$ (green), $C=-1$ (blue), $C=-2$ (pink))} 
	\label{fig_chi}%
\end{figure}

The Figs. \ref{fig_chi}a and \ref{fig_chi}b depict the profile of $\chi(x)$. Note from the figures that $\chi(x)$ has an antilump (for $C>0$) and lump (for $C<0$) profiles around $x=0$ with $\chi(0)=\arctan(C)$. Also we have 
$\chi(x\to\pm\infty)=\pi/2$. Let us first consider what happens with the reduction of small values of $|C|$.
For $C>0$ (Fig. \ref{fig_chi}a) the antilump reduces the minimum of $\chi$ until achieving a plateau around $\chi=0$ (north pole of the surface $S^2_{int}$).
 On the other hand, for $C<0$ (Fig. \ref{fig_chi}b) the lump grows the maximum of $\chi$ until achieving a plateau around $\chi=\pi$ (south pole of the surface $S^2_{int})$) that grows for smaller values of $C$.  Now, when $|C|$ grows, the difference between maxima and minima of $\chi$ is reduced, signaling to a lower influence of the $\chi$ field in the composition of the topological defect.  Taking together, we see that the fields $(\phi,\chi)$ form a composite topological defect that interpolate between two vacuum. 
 
 Let us call a soliton $K_{n,1}$ ($K_{n,-1}$) as the set of solutions $(\phi,\chi)$ where the $\phi$ field has a kink profile and the $\chi$ field has an antilump (lump) one, connecting the minima $(\phi,\chi)=(2n\pi,\chi)\to( 2(n+1)\pi,\chi)$. An antisoliton $\bar K_{n,1}$ ($\bar K_{ n,-1}$) would be 
 the $\phi$ field with an antikink profile,
  and the $\chi$ field with a antilump (lump) one, connecting the minima $(\phi,\chi)=(2(n+1)\pi,\chi)\to(2n\pi,\chi)$. 
 In particular, $K_{0,1}=(\phi,\chi)$ with $\phi(x)$ given by the Eq. (\ref{phix}), whereas $\bar K_{0,1}=(\phi_{\bar S},\chi)$, with
\be 
\phi_{\bar S}(x)=2\pi-4\arctan(e^x).
\ee
Also, since $\chi(-C,x)=-\chi(C,x)$ we can just restrict to $C>0$ and consider lump or antilump for our initial configurations.

\begin{figure}
    \includegraphics[{angle=0,width=8cm}]{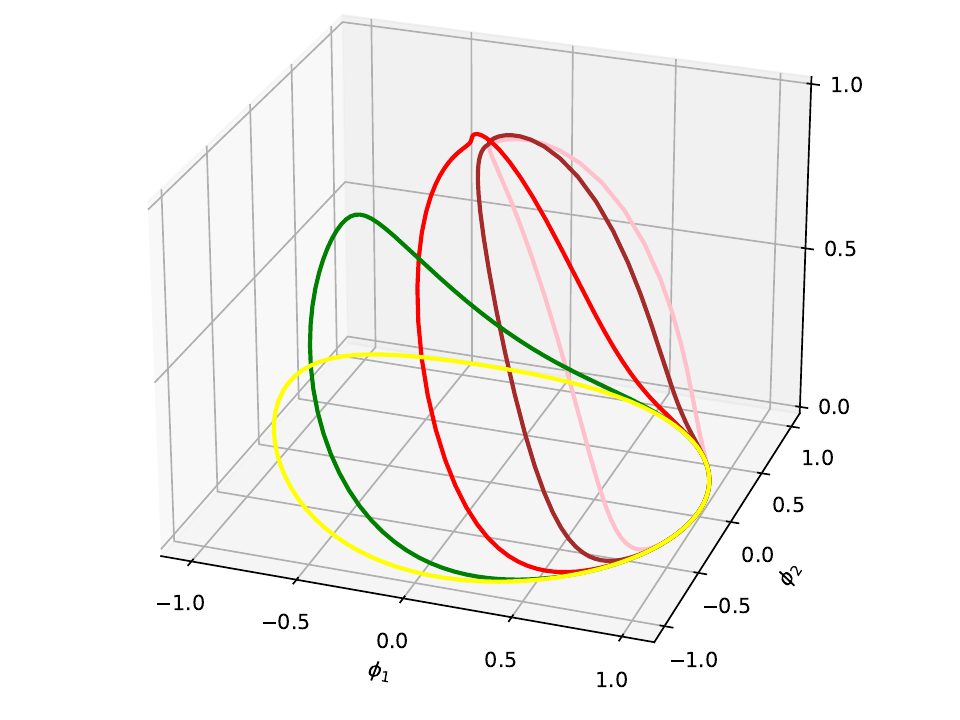}
    \includegraphics[{angle=0,width=8cm}]{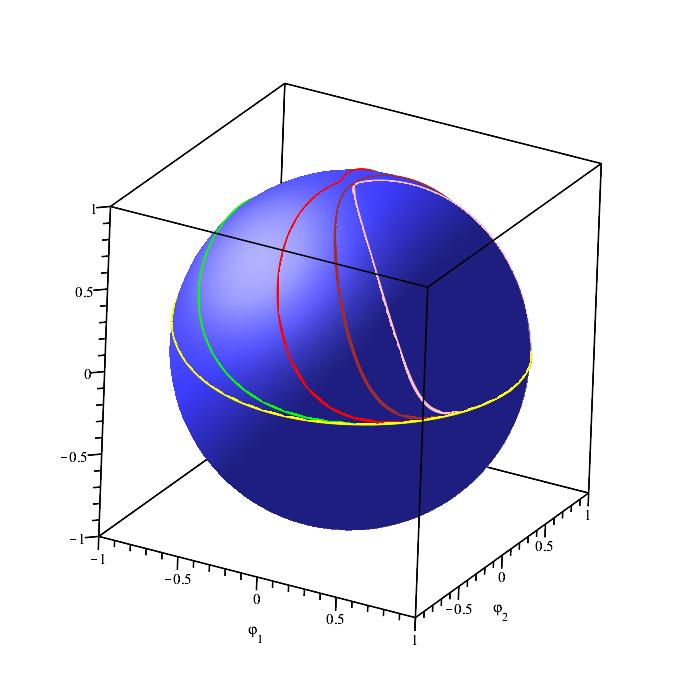}\\
    \includegraphics[{angle=0,width=5cm}]{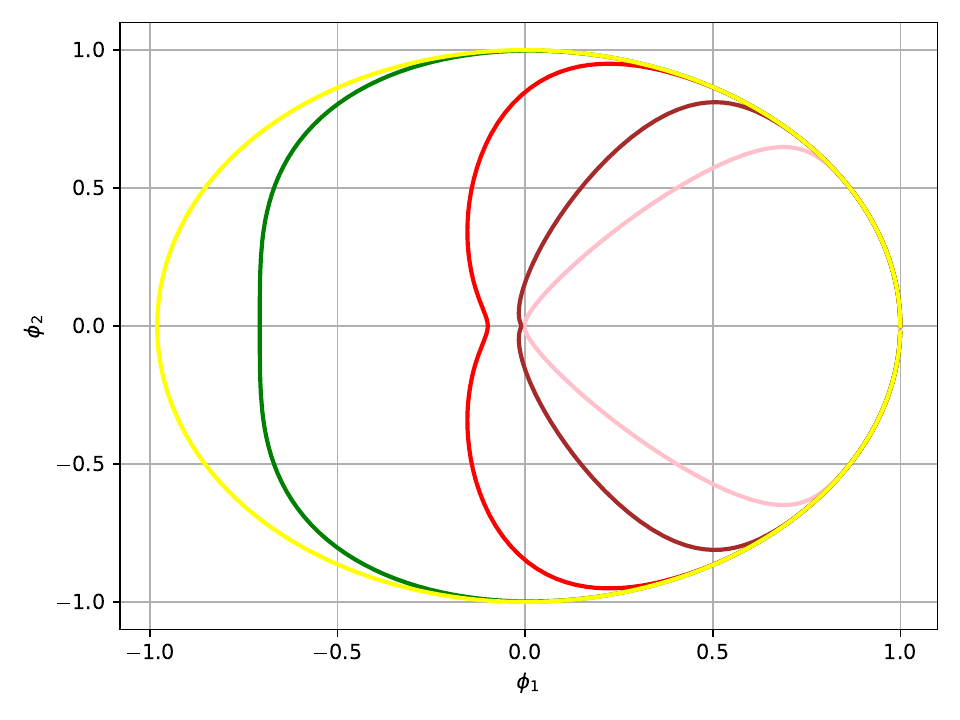}
    \includegraphics[{angle=0,width=5cm}]{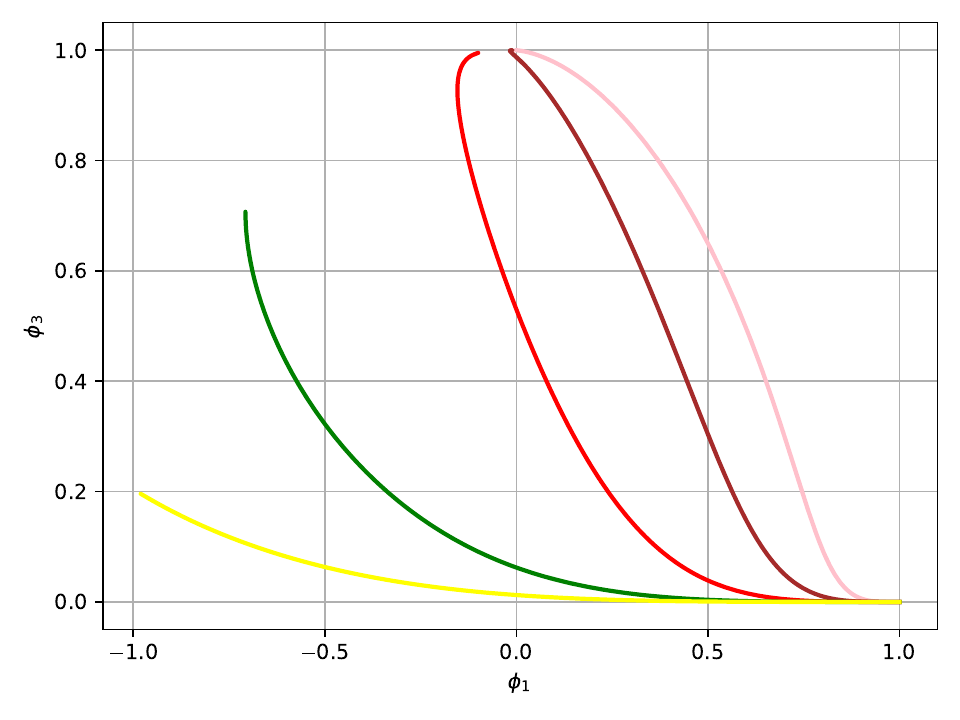}
    \includegraphics[{angle=0,width=5cm}]{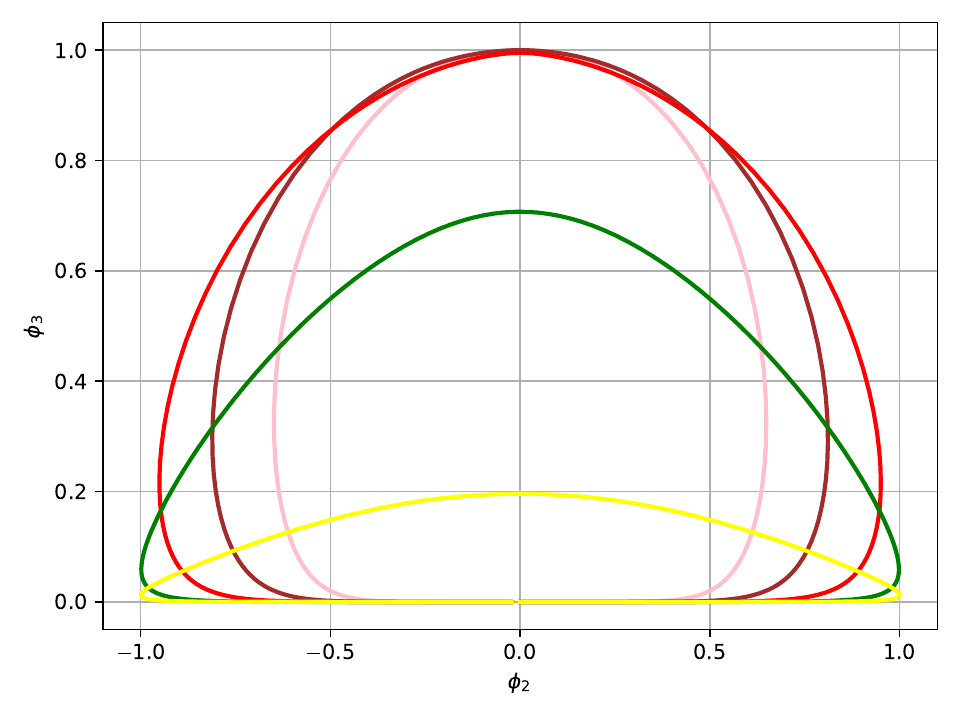} 
    \caption{ Soliton (antisoliton) solutions $K_{0,1}$ ($\bar K_{0,-1}$) in internal space for 
  $C=10^{-3}$ (pink), $C=10^{-2}$ (brown), $C=0.1$ (red), $C=1$ (green) and $C=5$ (yellow) a) the soliton (antisoliton) solutions are counterclockwise (clockwise) loops in internal space,  b) The soliton loops viewed in the $S_{int}^2$ internal space, c) solitons in phase space $\varphi_1\times\varphi_2$, d) solitons in phase space $\varphi_1\times\varphi_3$, e) solitons in phase space $\varphi_2\times\varphi_3$.}
	\label{fig_sphere_Cplus}%
\end{figure}

\begin{figure}
    \includegraphics[{angle=0,width=8cm}]{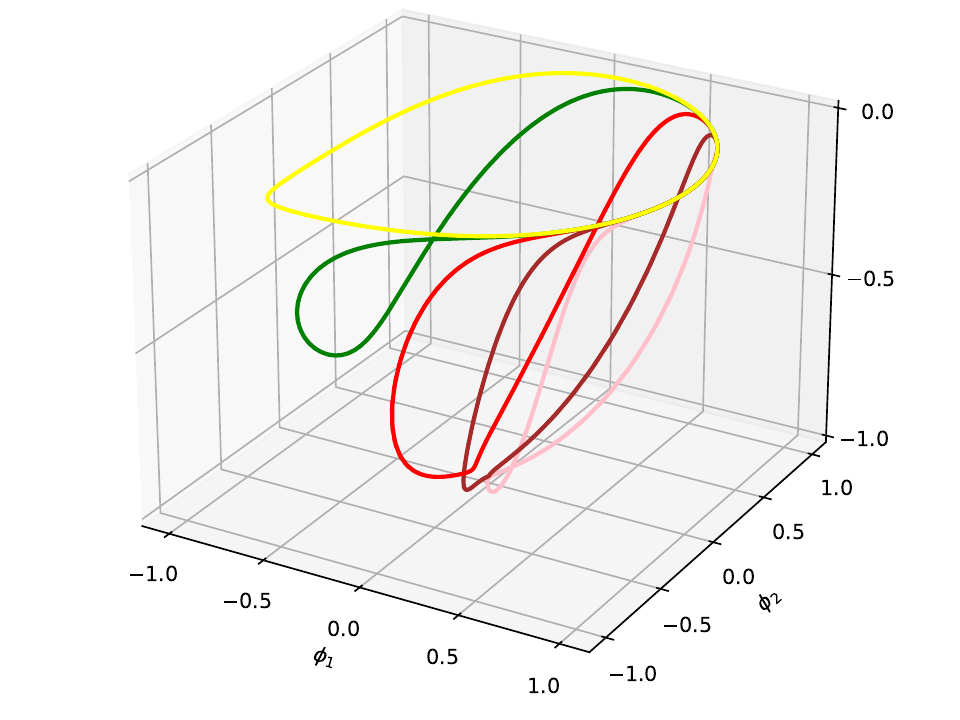}
    \includegraphics[{angle=0,width=8cm}]{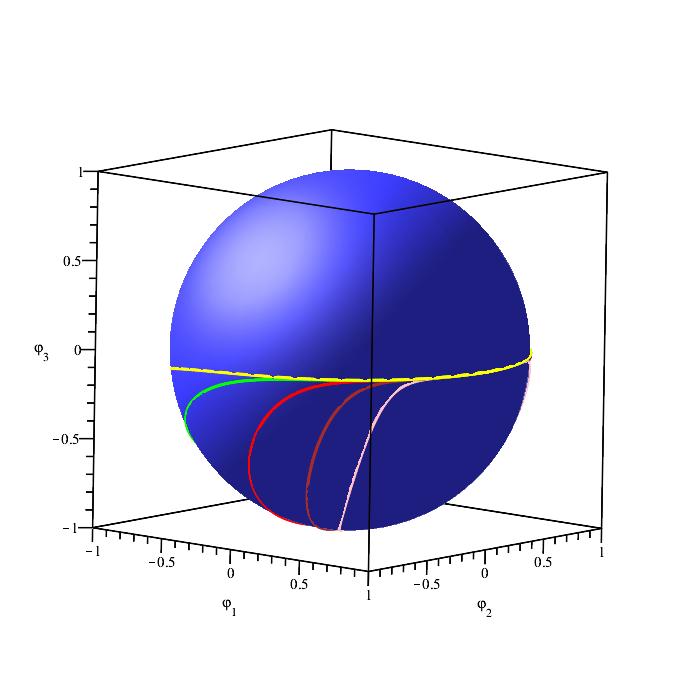}\\
    \includegraphics[{angle=0,width=5cm}]{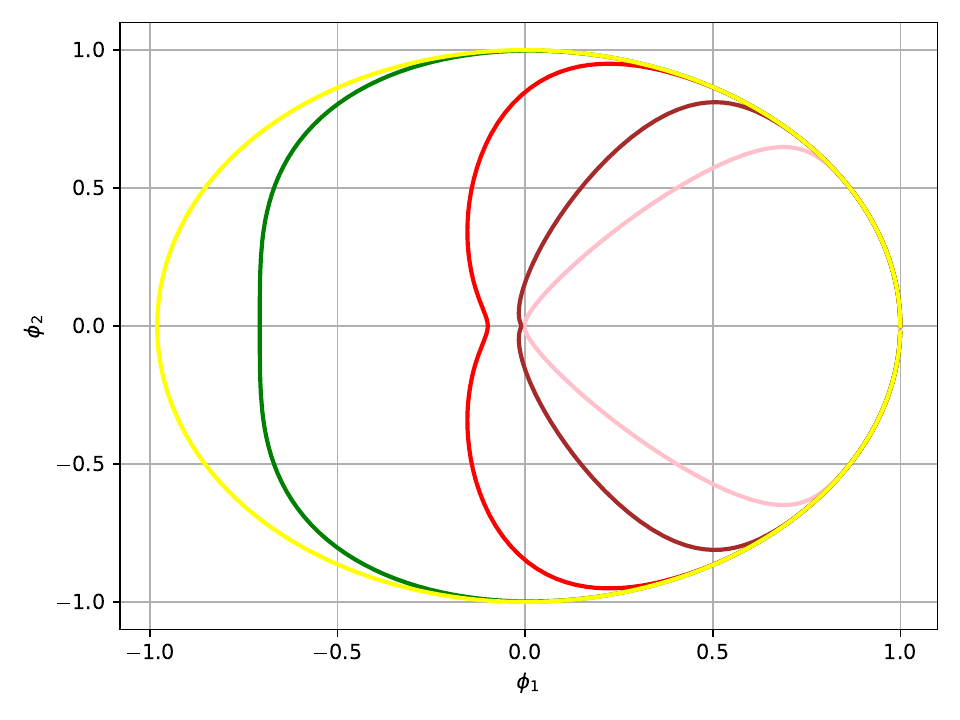}
    \includegraphics[{angle=0,width=5cm}]{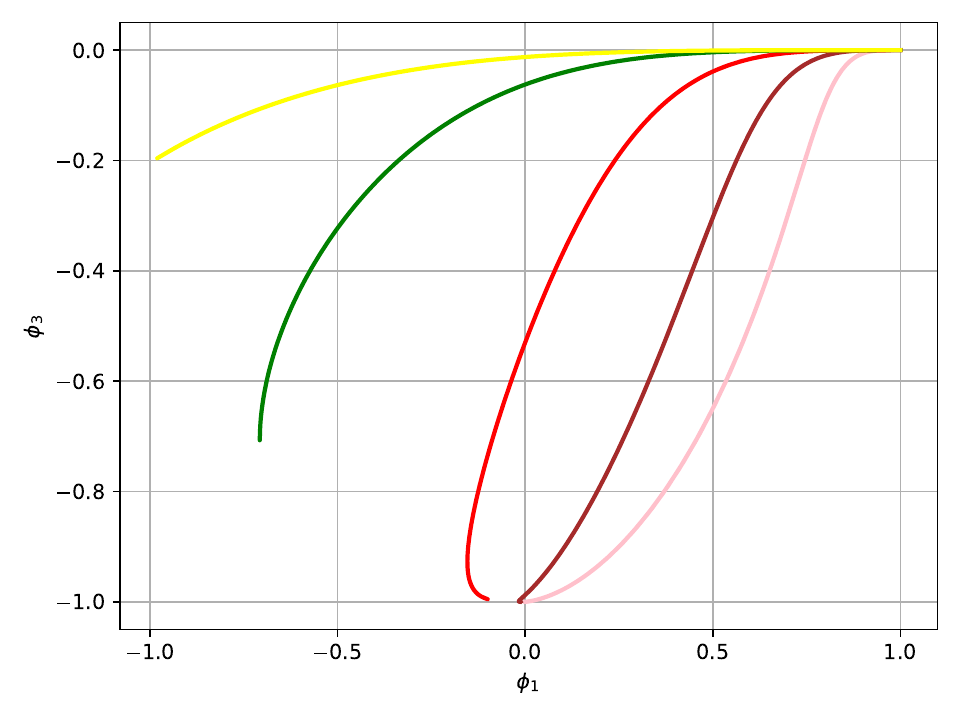}
    \includegraphics[{angle=0,width=5cm}]{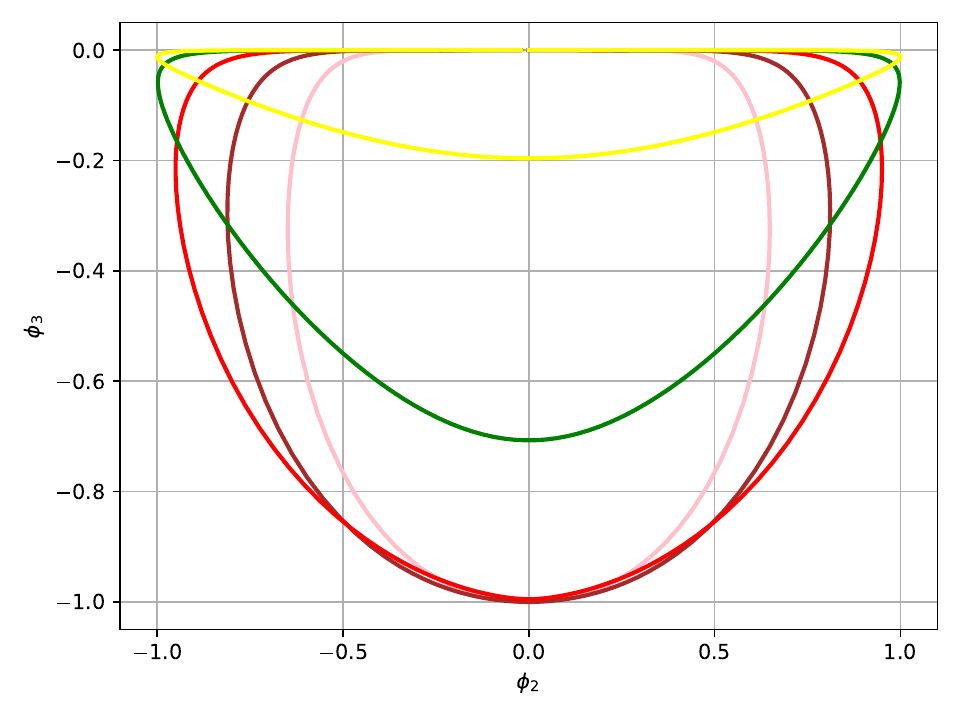} 
    \caption{ Soliton (antisoliton) solutions $K_{0,-1}$ ($\bar K_{0,-1}$) in internal space for 
  $C=10^{-3}$ (pink), $C=10^{-2}$ (brown), $C=0.1$ (red), $C=1$ (green) and $C=5$ (yellow) a) the soliton (antisoliton) solutions are counterclockwise (clockwise) loops in internal space, b) The soliton  loops viewed in the $S_{int}^2$ internal space, c) solitons in phase space $\varphi_1\times\varphi_2$, d) solitons in phase space $\varphi_1\times\varphi_3$, e) solitons in phase space $\varphi_2\times\varphi_3$. }
	\label{fig_sphere_Cminus}%
\end{figure}

The soliton configurations (corresponding to kink profile for $\phi$ and antilump profile for $\chi$) can be seen in the internal field space in the Figs. \ref{fig_sphere_Cplus}a-e (for $K_{0,1}$) and \ref{fig_sphere_Cminus}a-e (for $K_{0,-1}$). There one can see that the static soliton corresponds to a closed loop in the internal space $S_{int}^2$, and that solutions with different values of $C$ do not cross each other. Each soliton solution, when viewed from the north pole of $S_{int}^2$, is a  counterclockwise loop in internal space when $x$ goes from $-\infty$ to $+\infty$. We also showed in the Figs. \ref{fig_sphere_Cplus}b and \ref{fig_sphere_Cminus}b that the loops  are indeed in the internal space $S_{int}^2$.
From the same figures we note that static solutions for a particular choice of $C$ cover only part of the angular range of the hemisphere of $S_{int}^2$, since $\phi$ spread from $0$ to $2\pi$ whereas $\chi$ interpolates between $\pi/2$ and $\arctan (C)$. Details from the loops can be viewed in the two-dimensional slices depicted in the Figs. \ref{fig_sphere_Cplus}c-e (for $K_{0,1}$) and \ref{fig_sphere_Cminus}c-e (for $K_{0,-1})$. 
Note that the loops are in the $\varphi_1>0$ region (north hemisphere) for $K_{0,1}$ and $\varphi_1<0$ region (south hemisphere) for $K_{0,-1}$. Moreover, the projection of the loops in the 
$(\varphi_1,\varphi_2)$ plane are identical for fixed $C$ (compare the Figs. \ref{fig_sphere_Cplus}c and \ref{fig_sphere_Cminus}c). 

From the behavior of the solutions it is clear that, when viewed from the north pole of $S^2_{int}$, a counterclockwise loop means a soliton - corresponding to kink profile for $\phi$ and lump profile for $\chi$, whereas a clockwise loop means an antisoliton - corresponding to antikink profile for $\phi$ and lump profile for $\chi$. That is, the Fig. \ref{fig_sphere_Cplus} represent either $K_{0,1}$ or $\bar K_{0,1}$, just inverting the direction of circulation of the loops. The same applies to the \ref{fig_sphere_Cminus}: the loops represent either $K_{0,-1}$ (counterclockwise) or $\bar K_{0,-1}$ (clockwise). 

\begin{figure}
    \includegraphics[{angle=0,width=10cm}]{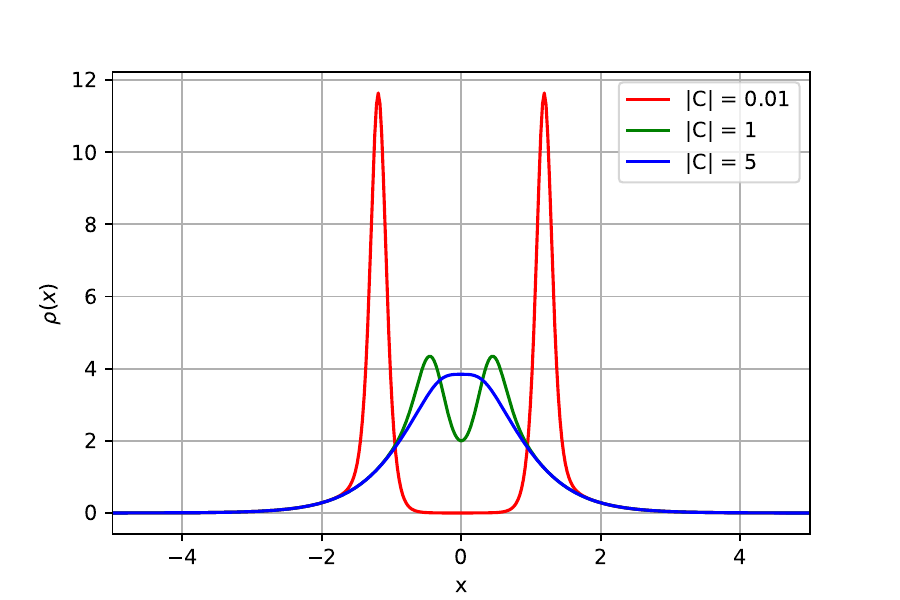}	
	\caption{ Energy density $\rho(x)$ of the soliton $K_{0,1}$ for some values of $C$.} 
	\label{fig_rho}%
\end{figure}

The Fig. \ref{fig_rho} shows the profile for the energy density $\rho(x)$ of the soliton $K_{0,1}$. Note from the figure that for $|C|<1$ the energy density is characterized by two peaks situated symmetrically from $x=0$. This signals the presence of an internal structure of the defect due to the changing of the scalar field $\chi(x)$, as discussed above. The peaks are more pronounced for lower values of $C$. Increasing $|C|$ the height of the peaks is reduced, and for $|C|\sim5$ there appears a central peak around $x=0$. Larger values of $C>5$ does not change the energy density significantly. 

\section{Scattering Results}

The geometric and potential coupling among the fields results in complex possibilities for scattering. Here we are interested in 
soliton-antisoliton scattering, that is, the
collision of $(\phi,\chi)$ topological defects such that dynamics can be viewed as kink-antikink for the $\phi$ field and two scenarios for the $\chi$ field. The first one is antilump-antilump, whereas the second one has a antilump-lump arrangement. We emphasize that due to the coupling, the collision occurs simultaneously for both fields.
\begin{figure}
	\includegraphics[{angle=0,width=8cm,height=6cm}]{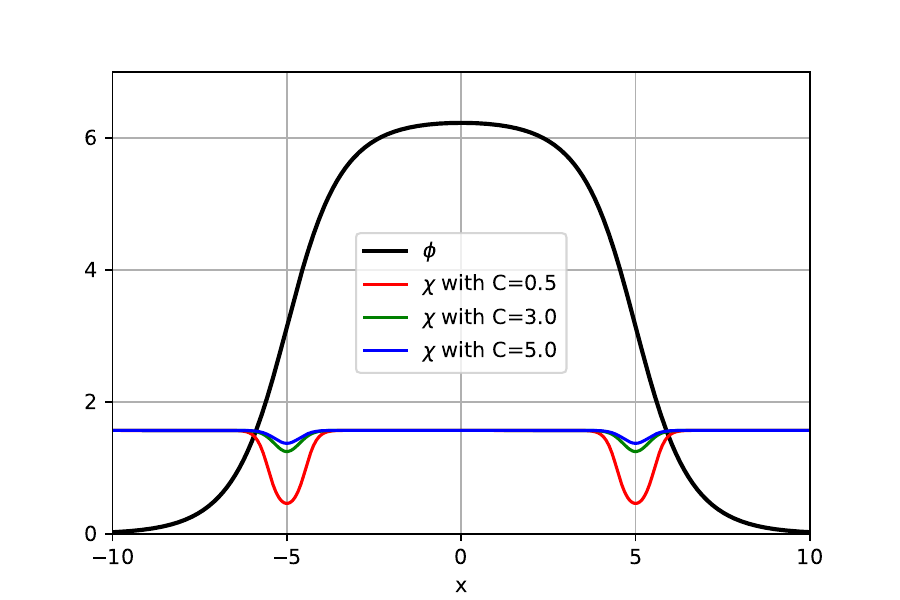}
	\includegraphics[{angle=0,width=8cm,height=6cm}]{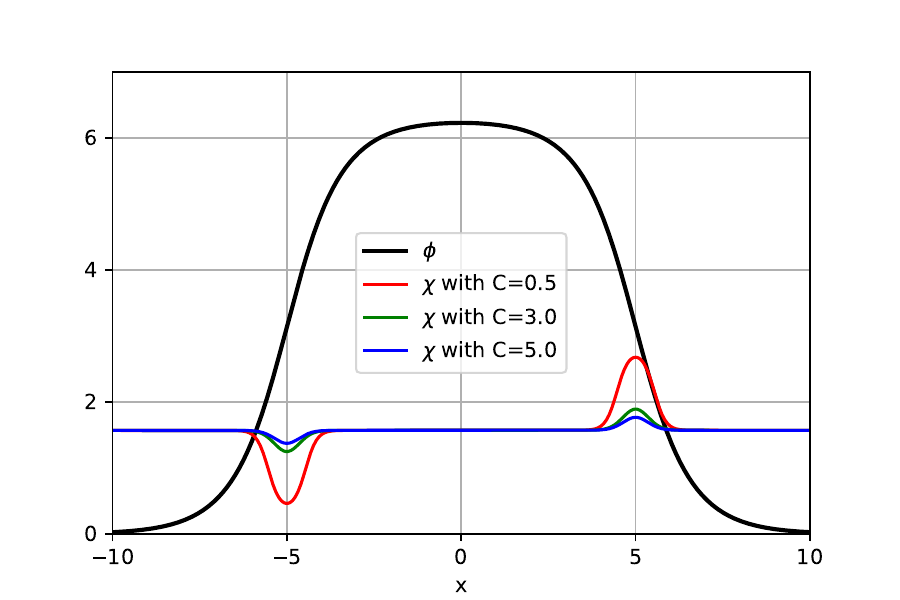}
	  \caption{Initial field configuration for soliton-antisoliton, meaning a) $K_{0,1}\bar K_{0,1}$, or  kink-antikink with antilump-antilump and b) $K_{0,1}\bar K_{0,-1}$ or kink-antikink with antilump-lump. For all figures, $\phi(x,t)$ (black) and $\chi(x)$ for $C=0.5$ (red), $C=3.0$ (green) and $C=10$ (blue). We fixed $x_0=5$} 
	\label{fig_kk}
\end{figure}

When viewed in the $S^2_{int}$ internal space, a soliton-antisoliton pair has roughly the same aspect of a loop as that observed for isolated soliton and antisoliton solution, with minor differences for $C\lesssim 0.1$. The main difference for the soliton-antisoliton pair is in the way the loop is circulated. To complete a map to the physical space from  $x\to -\infty$ to $x\to +\infty$, the loop must be circulated firstly in counterclockwise direction (then $x$ runs from  $x\to -\infty$ to $x\to 0$), and then in clockwise one (then $x$ runs from  $x\to -\infty$ to $x\to 0$).

In what follows, we describe the scattering process by solving the equations of motion in a box $-x_{max}<x<x_{max}$, where we consider $x_{max}=100$. Additionally, the partial derivatives with respect to $x$ were approximated using the five-point stencil with space step $\delta x=0.05$. The resulting set of equations was integrated via the fifth-order Runge-Kutta method with adaptive step size. We also have considered periodic boundary conditions. The initial configuration for some values of $C$ and $x_0=5$ is presented in the Fig. \ref{fig_kk}a for $K_{0,1}\bar K_{0,1}$ and Fig. \ref{fig_kk}b for $K_{0,1}\bar K_{0,-1}$. Note from the figure that $2x_0$ corresponds to the initial separation of the defects. The influence of $C$ was already discussed for an isolated defect and is the same for the initial condition of the two defects. 
\subsection{$K_{0,1}\bar K_{0,1}$ scattering}
In the following, we will discuss the main results for kink-antikink and antilump-antilump collision for the $\phi,\chi$ fields. In order to accomplish this, we used the following initial conditions to examine the scattering process
\begin{eqnarray}
\chi\left(x,0,x_0,v\right) &=& \chi_{L}\left( \gamma \left( x+ x_{0} - vt\right) \right) + \chi_{L}\left(\gamma \left(x-x_0 + vt\right) \right) - \frac{\pi}{2}\text{,}\\
\dot{\chi}\left(x,0,x_0,v\right) &=& \dot{\chi}_{L}\left( \gamma \left( x+ x_{0} - vt\right) \right) + \dot{\chi}_{L}\left(\gamma \left(x-x_0 + vt\right) \right) \text{,}
\end{eqnarray}
and
\begin{eqnarray}
\phi \left(x,0,x_0,v\right) &=&\phi_{K}\left( \gamma \left( x+ x_{0} - vt\right) \right) +\phi_{\bar{K}}\left( \gamma \left(
x- x_{0} + vt\right) \right) -2\pi \text{,}\\
\dot{\phi} \left(x,0,x_0,v\right) &=&\dot{\phi}_{K}\left( \gamma \left( x+ x_{0} - vt\right) \right) +\dot{\phi}_{\bar{K}}\left( \gamma \left(x- x_{0} + vt\right) \right) \text{,}
\end{eqnarray}%
where these expressions correspond to the static solutions subjected to the Lorentz factor $\gamma=(1-v^2)^{-1/2}$. In the following, we will discuss the main results of the collision between scalar fields for $x_0=3$. 

\begin{figure*}[!ht]
\begin{center}
  \centering
    \subfigure[ ]{\includegraphics[{angle=0,height=3cm,width=12cm}]{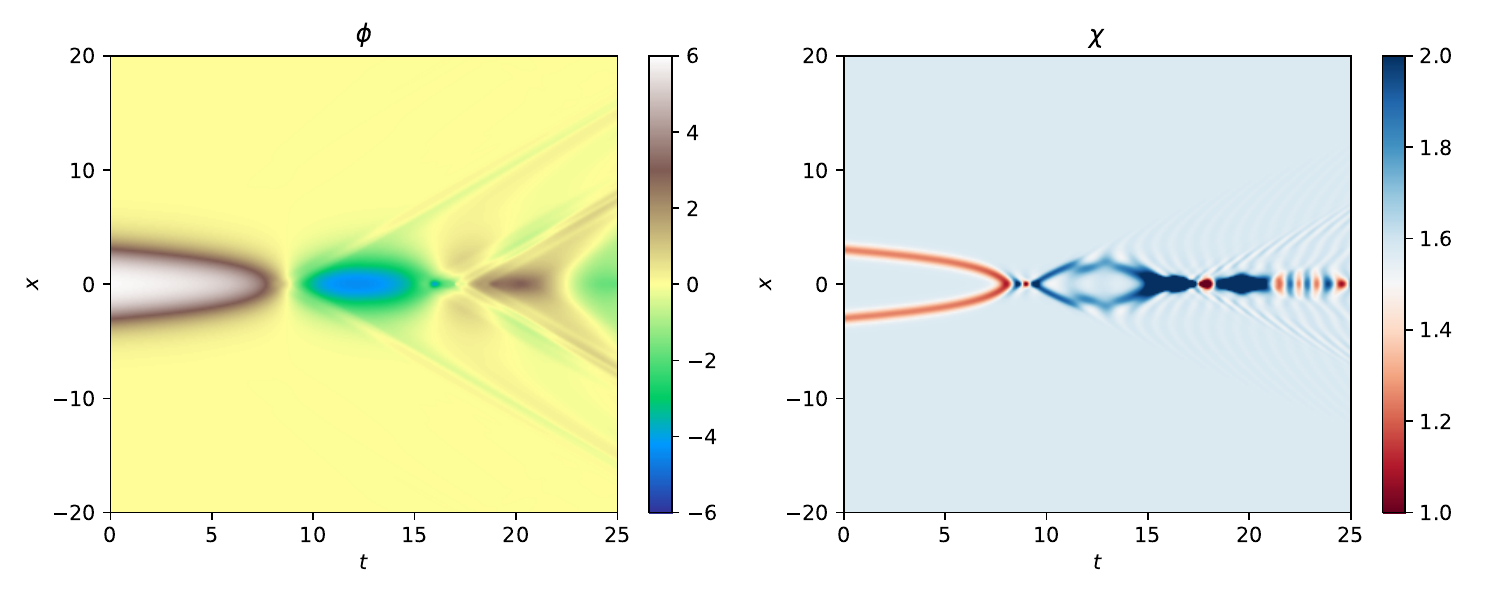}\label{v015_c3}}
    \subfigure[ ]{\includegraphics[{angle=0,height=3cm,width=12cm}]{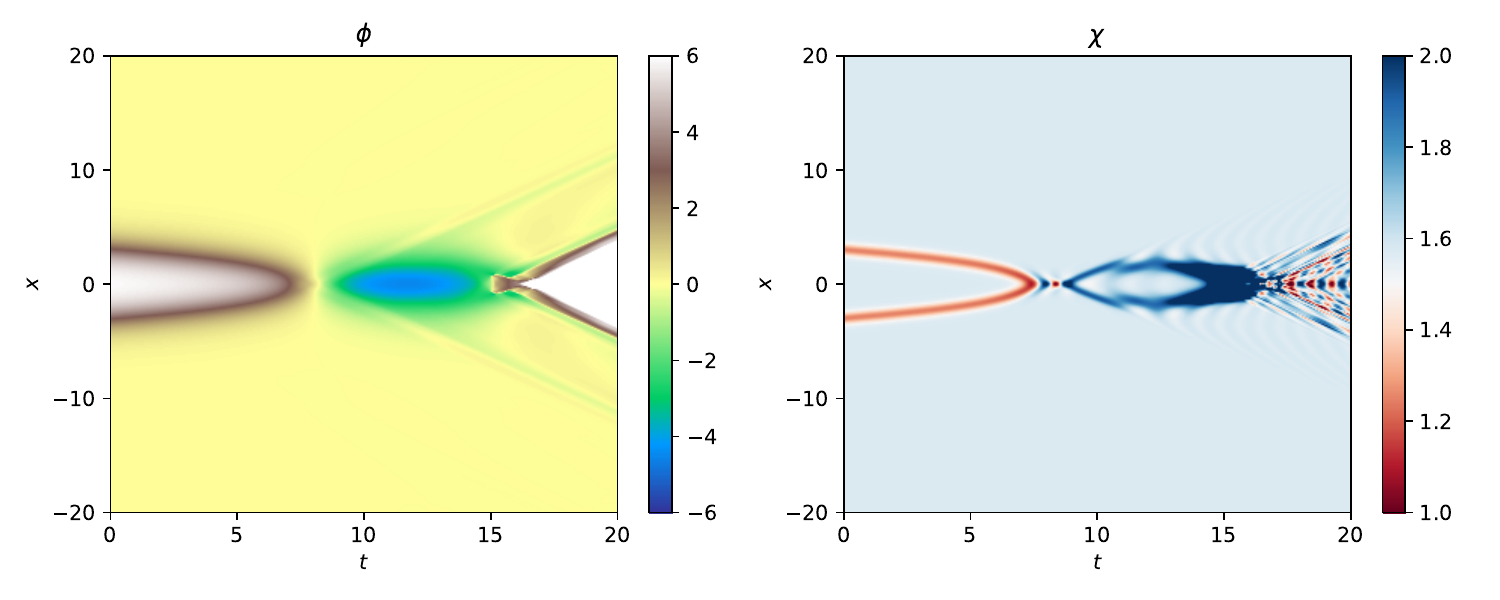}\label{v018_c3}}
    \subfigure[ ]{\includegraphics[{angle=0,height=3cm,width=12cm}]{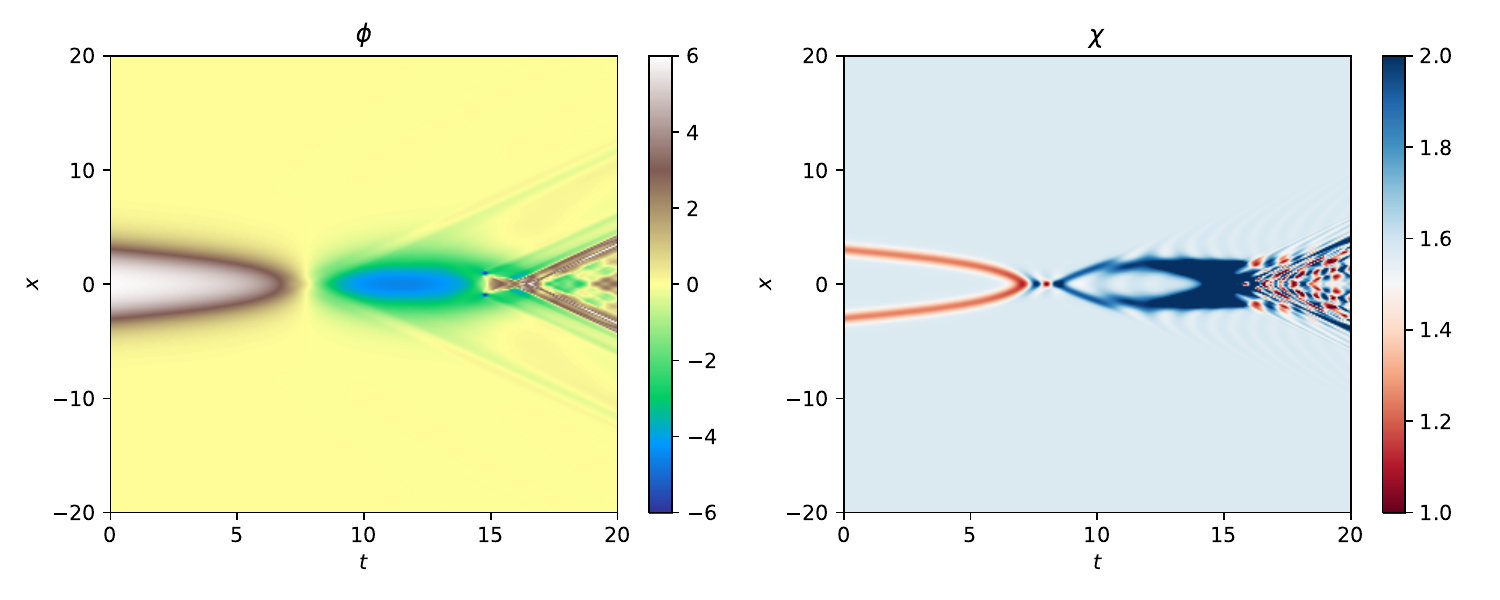}\label{v02_c3}}
    \subfigure[ ]{\includegraphics[{angle=0,height=3cm,width=12cm}]{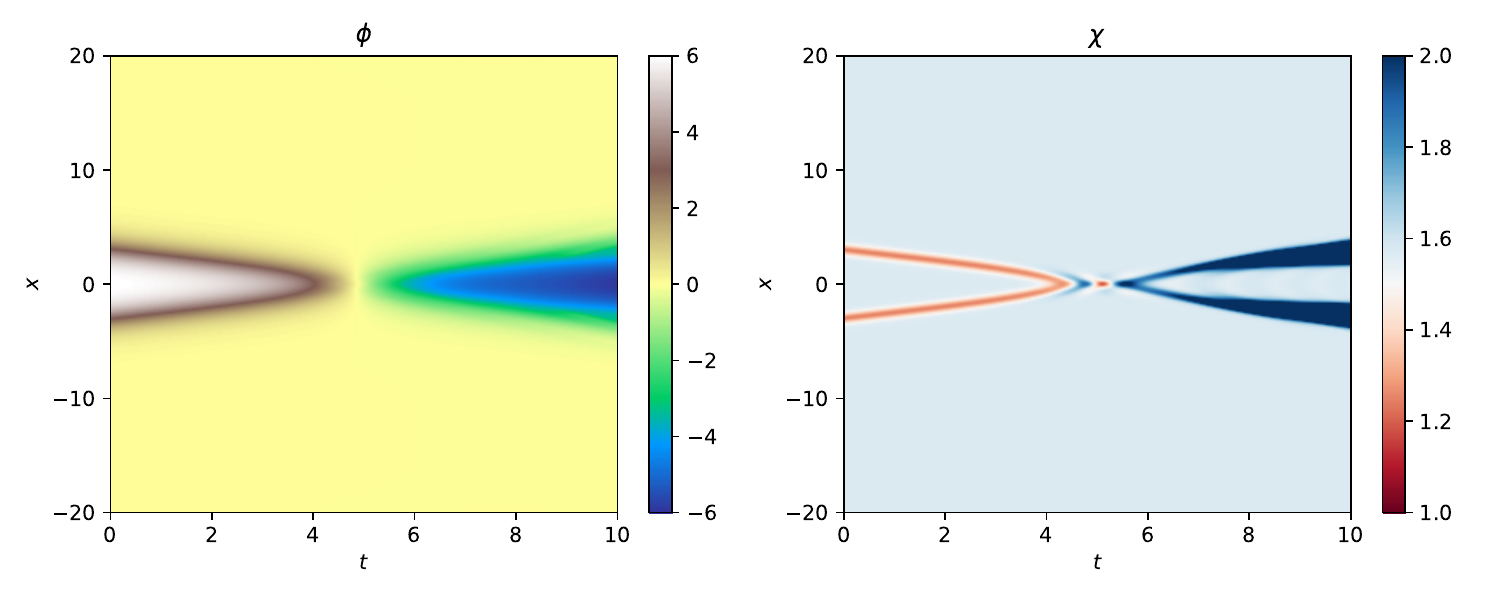}\label{v05_c3}}
        \caption{$K_{0,1}\bar K_{0,1}$ scattering. Fields (left) $\phi(x,t)$ and (right) $\chi(x,t)$ for a) $v=0.15$, b) $v=0.18$, c) $v=0.20$ and d) $v=0.50$. We fixed $C=3$.} 
	\label{fig_scat}
\end{center}
\end{figure*}

In the Fig. \ref{fig_scat} we highlight the  scattering results for $C=3$ for four small values of $v$. For $v=0.15$ (Fig. \ref{fig_scat}a) we noticed a strong emission of radiation after the collision, mainly for the $\phi$ field. It should be noted that a process of vacuum exchange starts after the kink-antikink pair collides and continues until the pair is annihilated. In the $\chi$ field, antilump-antilump collisions can be observed. The pair seems to separate indefinitely, but they lack the necessary force to overcome their mutual attraction. Finally, they collide a few more times, forming a bion around $x=0$ before annihilating each other. As the initial velocity increases, the scattering output changes significantly, particularly in the $\phi$ field. For instance, in Fig. \ref{fig_scat}b for $v=0.18$, the time evolution of the kink-antikink pair exhibits two-bounce behavior. Note the creation of a pair after the kinks collide twice. Nevertheless, antilump-antilump scattering still leads to complete annihilation. The Fig. \ref{fig_scat}c  depicts the results for the initial velocity of $v=0.20$. For the $\phi$ field one can see central oscillations, as well as two kink-antikink pairs scattering out in opposite directions. The antilump-antilump still exhibits annihilation behavior, in a way similar to observed in the Fig. \ref{fig_scat}b.
Now, one can see that the behavior of the $\chi$ field changes as the initial velocity increases further. This can be seen, for instance, for $v=0.50$ in the Fig. \ref{fig_scat}d. There note that the incoming $\chi$ antilump-antilump pair collides only once before dispersing like two lump-lump pairs. For the $\phi$ field, we have the abrupt change of the scalar field at the center of mass. Then, the scattering gives roughly a simple change of topological sector, that could be  represented by 
$K_{0,1}+\bar K_{0,1} \to \bar K_{-1,-1}+ K_{-1,-1}$. However, we stress that this is not an elastic scattering, since there is some radiation emission. 

\begin{figure*}[!ht]
\begin{center}
  \centering
    \subfigure[ ]{\includegraphics[{angle=0,height=3cm,width=12cm}]{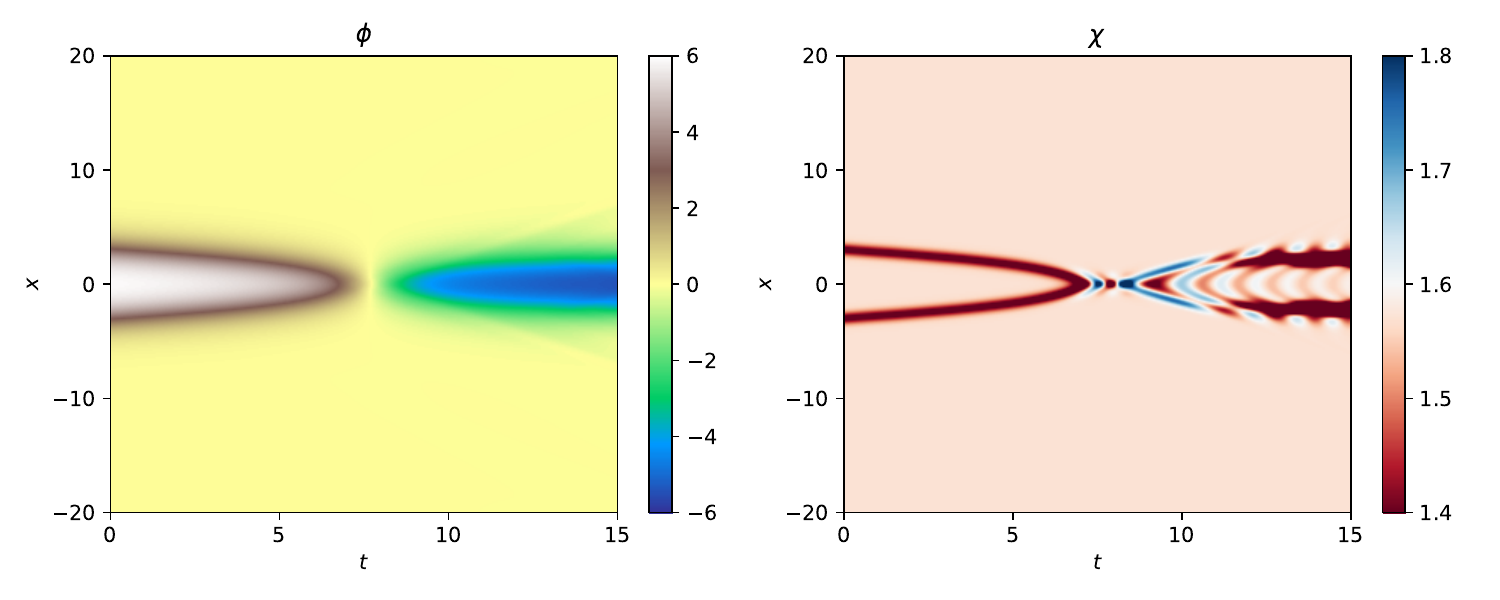}\label{v02_c5}}
    \subfigure[ ]{\includegraphics[{angle=0,height=3cm,width=12cm}]{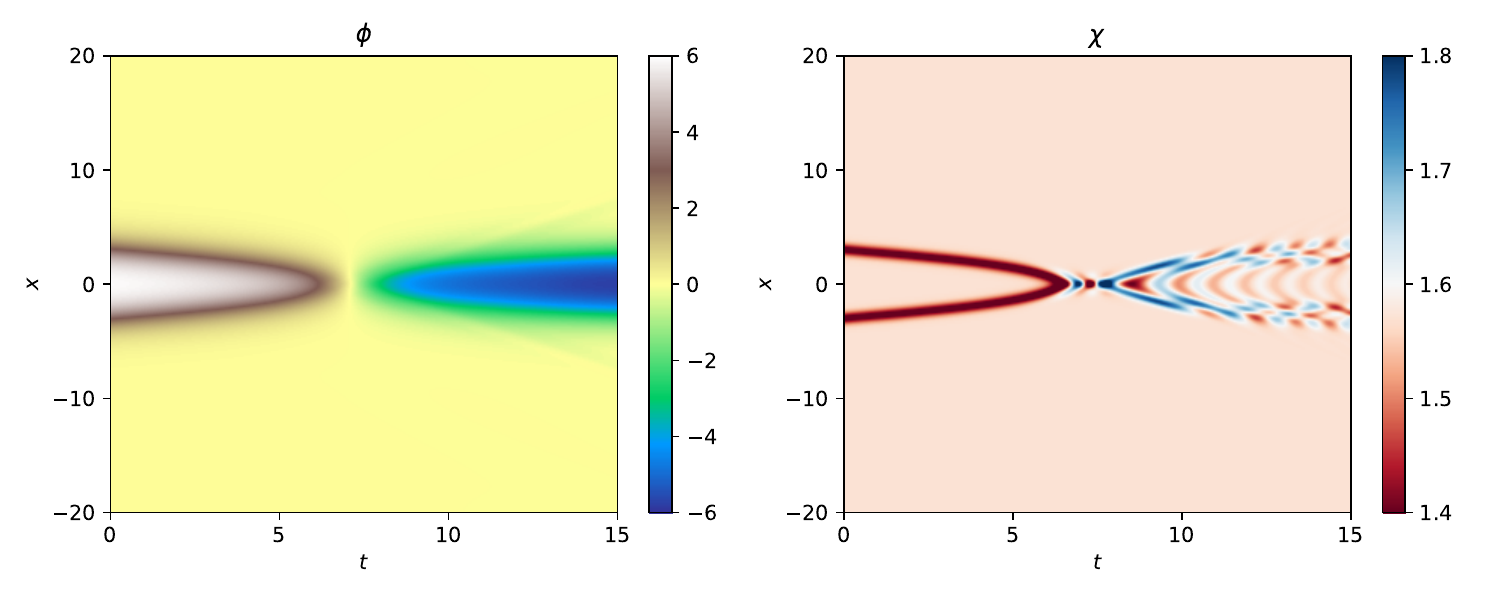}\label{v024_c5}}
    \subfigure[ ]{\includegraphics[{angle=0,height=3cm,width=12cm}]{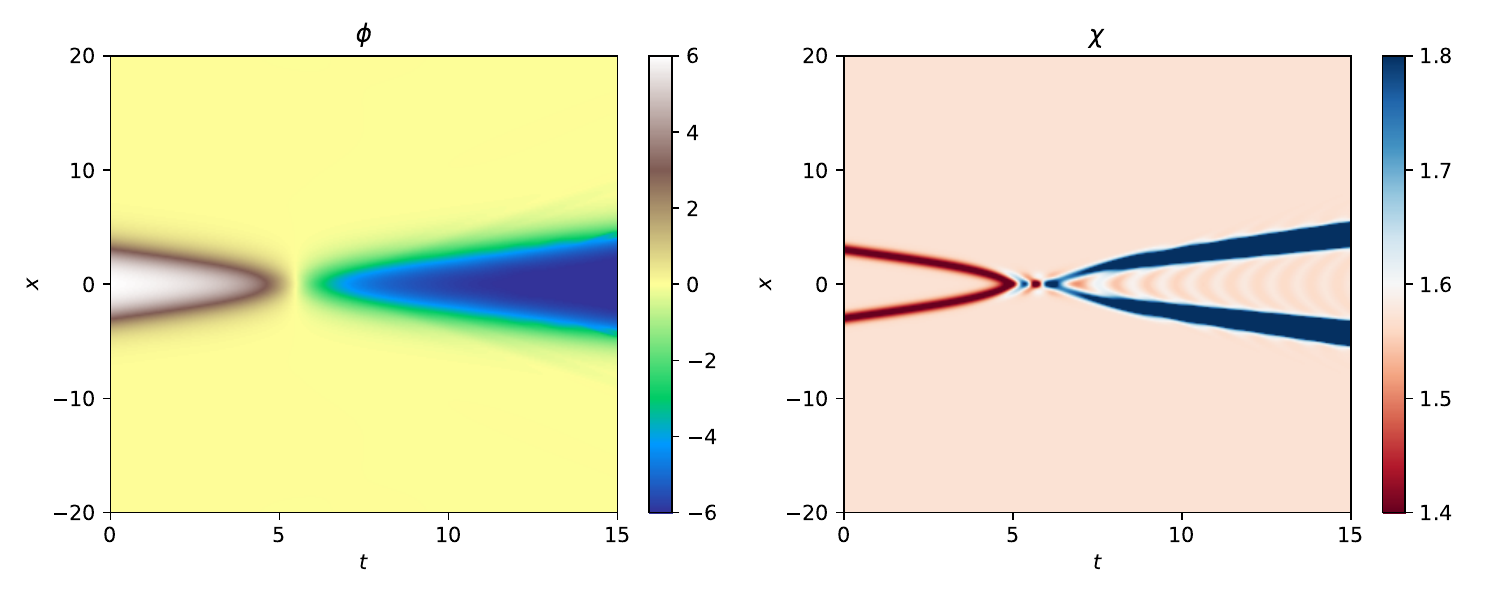}\label{v04_c5}}
        \caption{$K_{0,1}\bar K_{0,1}$ scattering. Fields (left) $\phi(x,t)$ and (right) $\chi(x,t)$ for a) $v=0.20$, (b) $v=0.24$,  (c) $v=0.40$. We fixed $C=5$.} 
	\label{fig_scat2}
\end{center}
\end{figure*}

We also considered the case with $C=5$. In the Fig.~\ref{fig_scat2}, we present some results of the collision between $v=0.20$ and $0.40$. For all cases, kink-antikink scattering for the $\phi$ field reveals a one-bounce collision with vacuum exchange. On the other hand, the $\chi$ field shows different final results: for $v=0.2$ (Fig.~\ref{fig_scat2}a),the antilump-antilump character of the field is maintained,  while for $v=0.4$ (Fig.~\ref{fig_scat2}c), the antilump-antilump profile for the $\chi$ field is changed to a lump-lump one. The Fig.~\ref{fig_scat2}b for $v=0.24$ depicts the presence of a transition region for an intermediate initial velocity. In this case, the $\phi$ field exhibits its typical one-bounce behavior, while the $\chi$ field shows a almost annihilation between the pair.

\begin{figure*}[!ht]
\begin{center}
  \centering
    \subfigure[ ]{\includegraphics[{angle=0,height=6cm,width=17cm}]{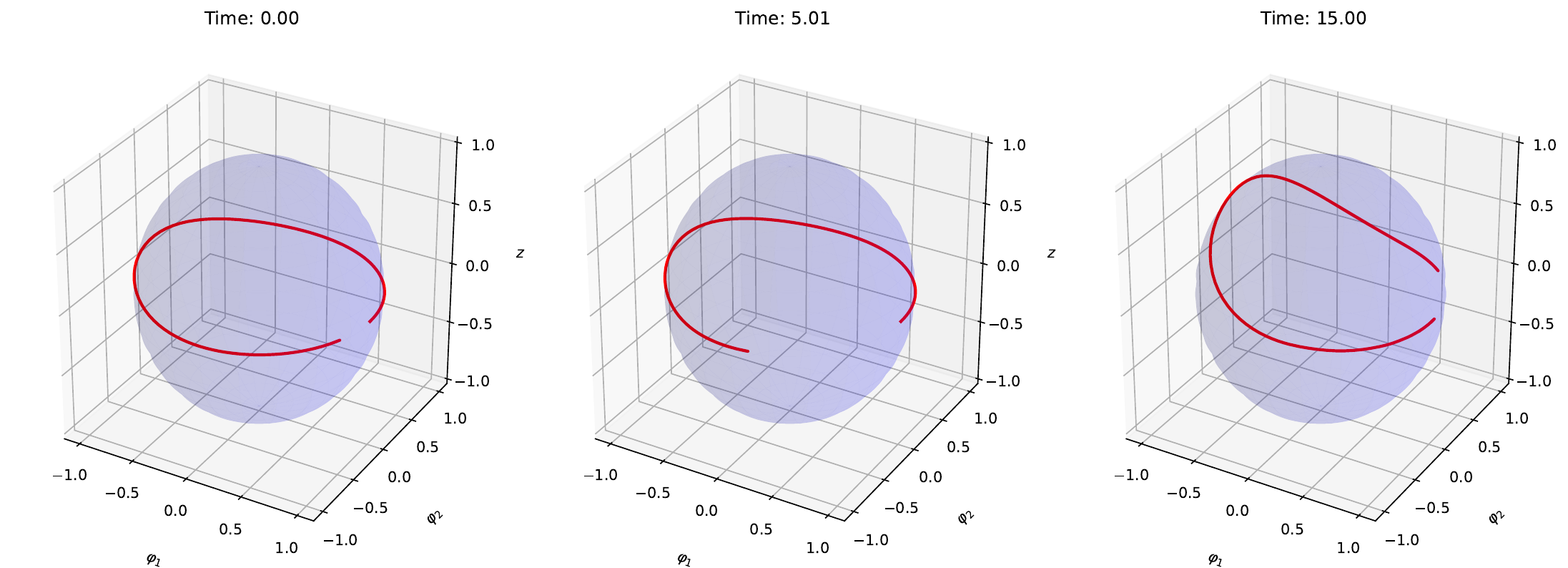}\label{sp_v02_c5}}
    \subfigure[ ]{\includegraphics[{angle=0,height=6cm,width=17cm}]{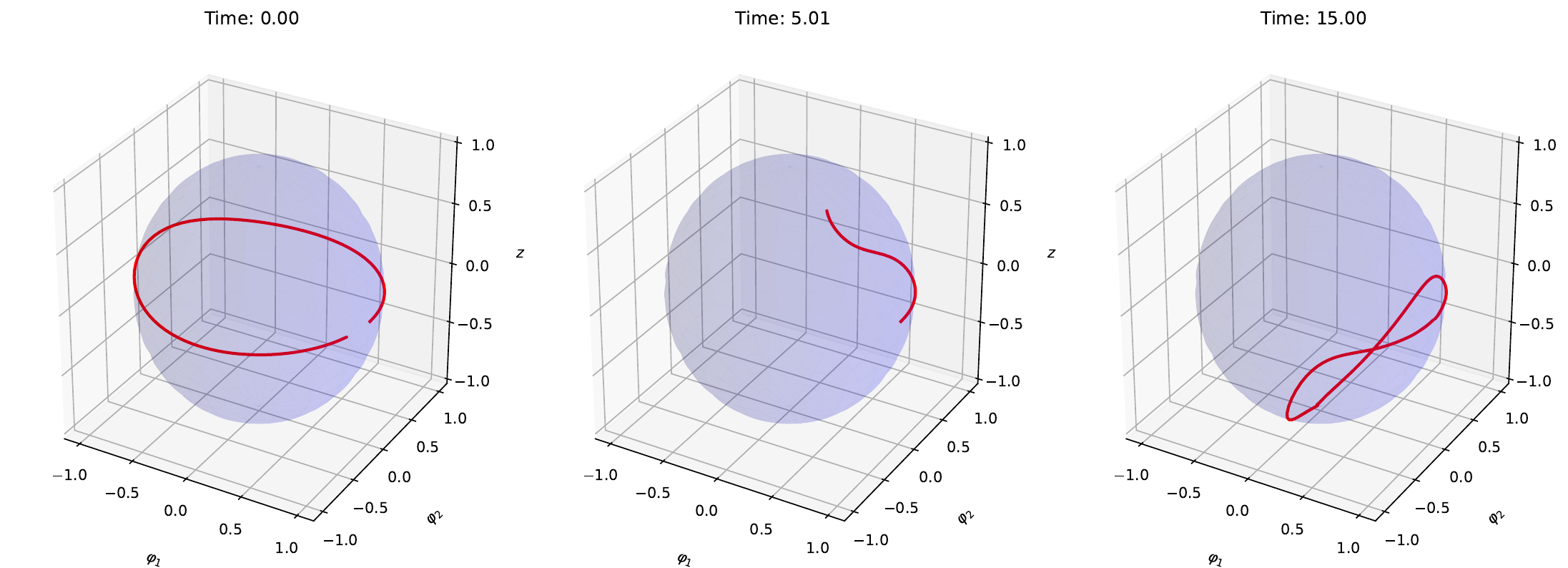}\label{sp_v04_c5}}
        \caption{$K_{0,1}\bar K_{0,1}$ scattering. Solutions in internal space  for a) $v=0.20$ and b) $v=0.40$. We fixed $C=5$. Videos from the scattering can be seen, respectively, in the supplementary material video1.mp4 and video2.mp4.} 
	\label{fig_sphere1}  
\end{center}
\end{figure*}

Note also that the radiation released during the process of scattering between the fields decreases as the value of the $C$ parameter rises (compare the Fig. \ref{fig_scat2}a-c with the Fig. \ref{fig_scat}a-d ). In particular, for large and finite $C$, the behavior for the $\phi$ field appears to approach that of the sine-Gordon model, without significant radiation emission. In contrast, for the $\chi$ field, the behavior is more complex and depends on the initial velocity $v$.

It is instructive to see the scattering process also in the internal field space. This is done for $C=5$ in the Fig. \ref{fig_sphere1}a-c for the same parameters of the Fig. \ref{fig_scat2}a and \ref{fig_scat2}c.  The Fig. \ref{fig_sphere1}a, for $v=0.20$, shows that the initial loop in internal space is partially suppressed, turned to a string,  deformed with time, ending in a loop more to the north. This is in accord with the Fig.\ref{fig_scat2}a, where the $\chi$ field maintain its profile antilump-antilump, with the reduction of $\chi$ after the collision. 
The Fig. \ref{fig_sphere1}b, for $v=0.40$, shows that the final loop is turned to the south hemisphere, compatible to the changing antilump-antilump $\to$ lump-lump for the $\chi$ field, as remarked before (see the Fig. \ref{fig_scat2}c). 

\begin{figure}
    \includegraphics[{angle=0,width=8cm,height=5cm}]{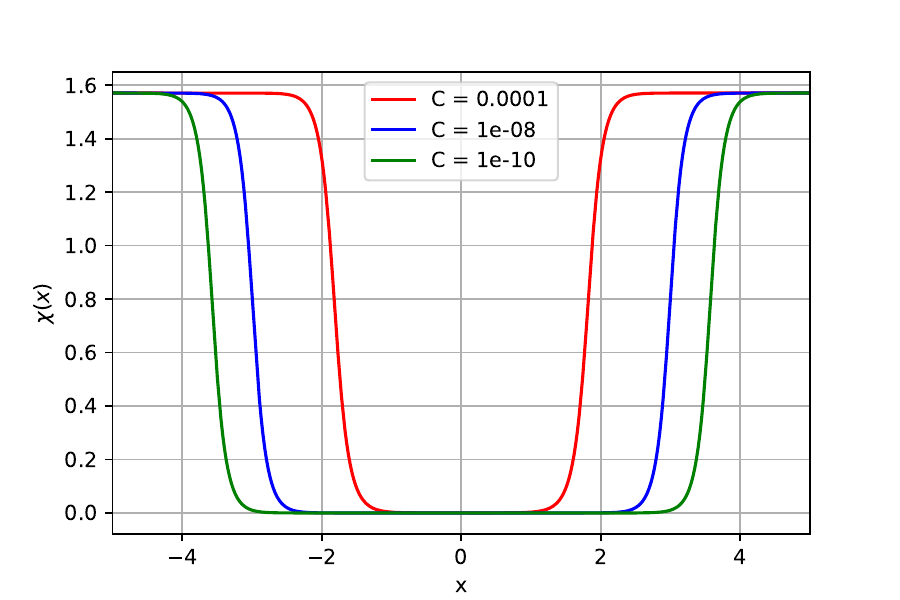}
    \includegraphics[{angle=0,width=8cm,height=5cm}]{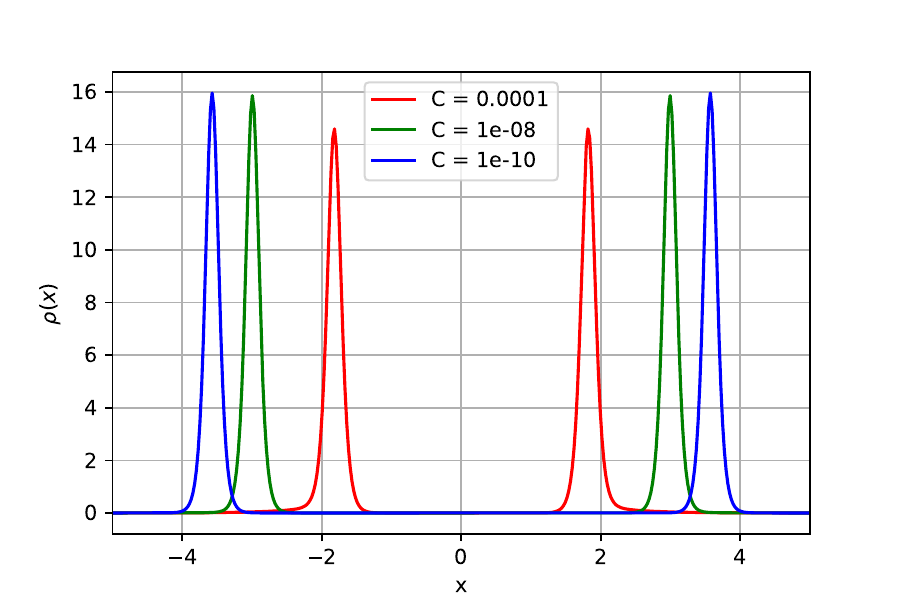}
    \caption{Soliton $K_{0,1}$: a) antilump scalar field $\chi(x)$, b) energy density $\rho(x)$ for $C=10^{-4}$ (red), $C=10^{-8}$ (blue) and $C=10^{-10}$ (green).}
    \label{fig_small}
\end{figure}

The numerical investigation for small values of $C$ is a hard task due to numerical instabilities when the $\chi$ field approaches zero. The influence of $C\ll1$ in the antilump profile for $\chi$ is presented in the Fig. \ref{fig_small}a. Note from the figure that $\chi$ is close to zero up to a finite distance from the center of mass, when the field grows until the vacuum $\chi=\pi/2$. Note that the smaller is $C$, the larger is this distance.
 This is in accord with an energy density profile characterized by two separated peaks, as shown in the 
\ref{fig_small}b (see also the Fig. \ref{fig_rho} for $|C|=0.01$). There we see that the $\chi$ field is the only one responsible for the changing of the energy density. Also, the limit $C\to 0$ is not the appropriate limit for searching for the model to recover the integrable sine-Gordon model. On the contrary, it is the 
$C\to\infty$ limit that is expected for suppressing the emitted radiation, as shown in the results for $C=3$ and $C=5$. Even for  these not too large values of $C$, the presence of the $\chi$ field still leads to significant emission of radiation for small velocities, as we saw in the Figs. \ref{fig_scat} and \ref{fig_scat2}. 

\subsection{$K_{0,1}\bar K_{0,-1}$
scattering}
Now, we will show the main outputs of the scattering process for kink-antikink and antilump-lump for the $\phi,\chi$ fields. We use the following initial conditions 
\begin{eqnarray}
\chi\left(x,0,x_0,v\right) &=& \chi_{L}\left( \gamma \left( x+ x_{0} - vt\right) \right) + \chi_{\bar{L}}\left(\gamma \left(x-x_0 + vt\right) \right) + \frac{\pi}{2} \text{,}\\
\dot{\chi}\left(x,0,x_0,v\right) &=& \dot{\chi}_{L}\left( \gamma \left( x+ x_{0} - vt\right) \right) + \dot{\chi}_{\bar{L}}\left(\gamma \left(x-x_0 + vt\right) \right) \text{,}
\end{eqnarray}
and
\begin{eqnarray}
\phi \left(x,0,x_0,v\right) &=&\phi_{K}\left( \gamma \left( x+ x_{0} - vt\right) \right) +\phi_{\bar{K}}\left( \gamma \left(
x- x_{0} + vt\right) \right) -2\pi \text{,}\\
\dot{\phi} \left(x,0,x_0,v\right) &=&\dot{\phi}_{K}\left( \gamma \left( x+ x_{0} - vt\right) \right) +\dot{\phi}_{\bar{K}}\left( \gamma \left(x- x_{0} + vt\right) \right) \text{,}
\end{eqnarray}%

\begin{figure}	\includegraphics[{angle=0,height=3cm,width=12cm}]{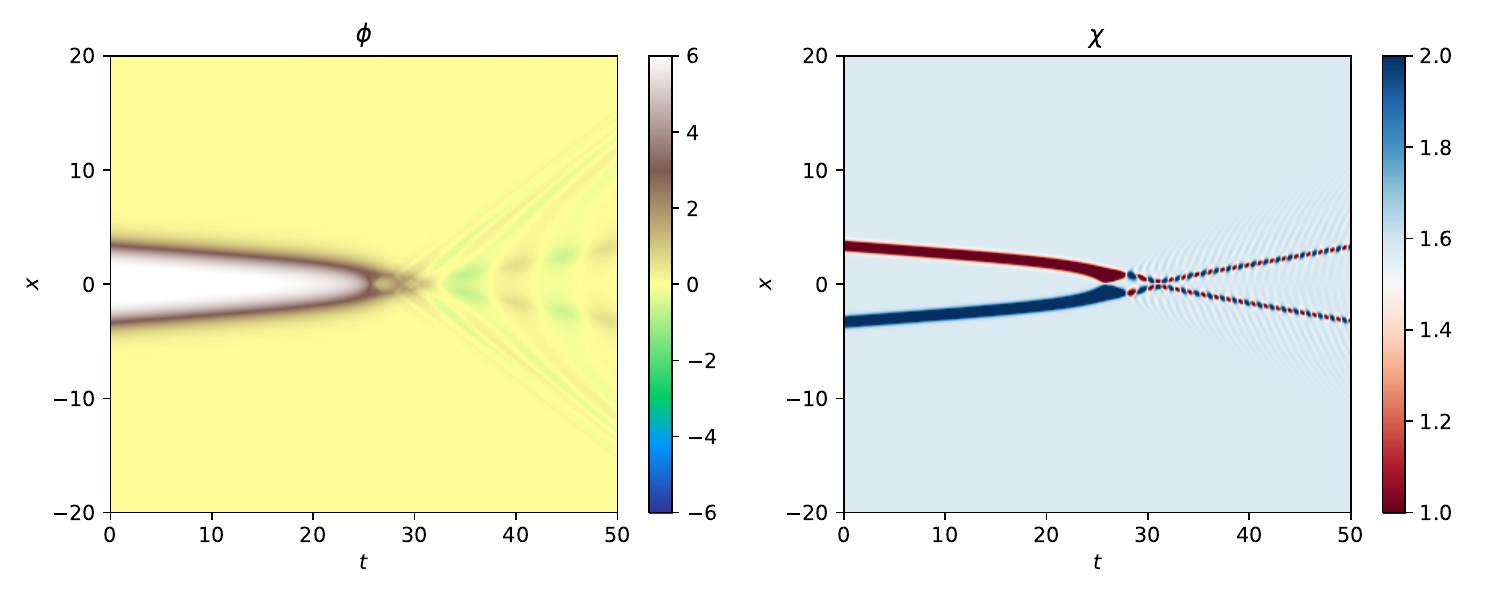}
\\
\includegraphics[{angle=0,height=3cm,width=12cm}]{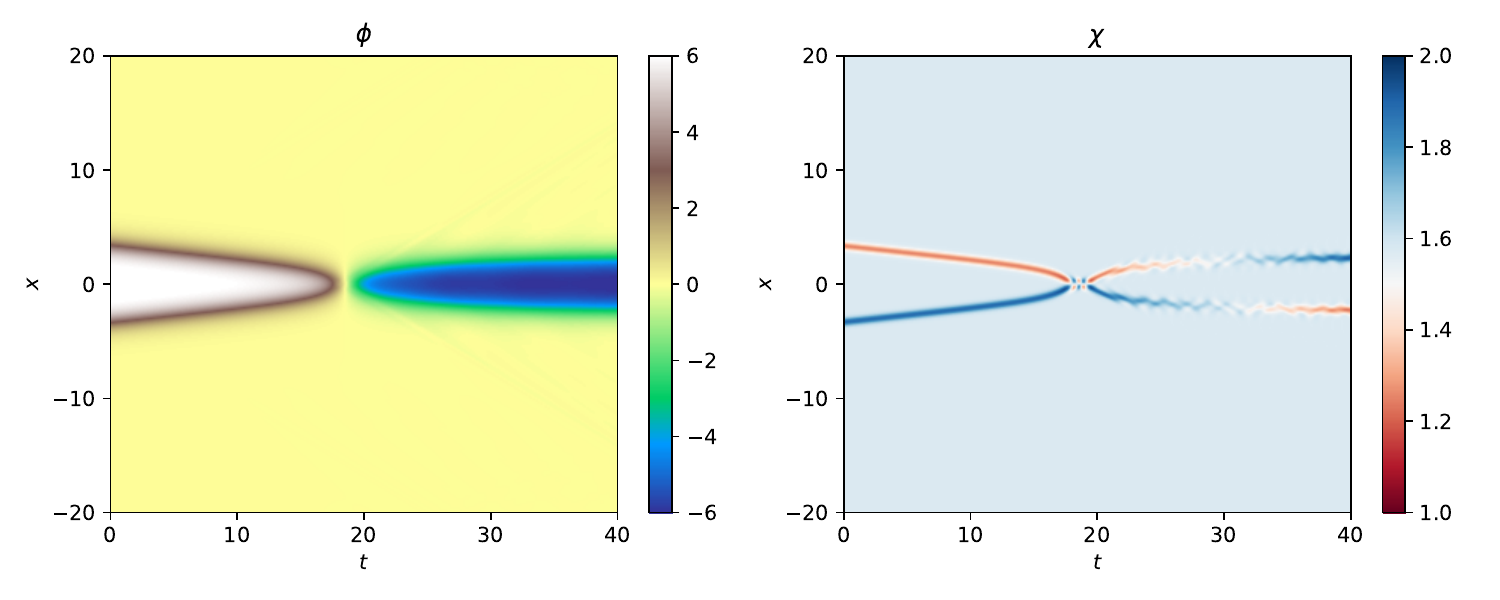}
\\
\includegraphics[{angle=0,height=3cm,width=12cm}]{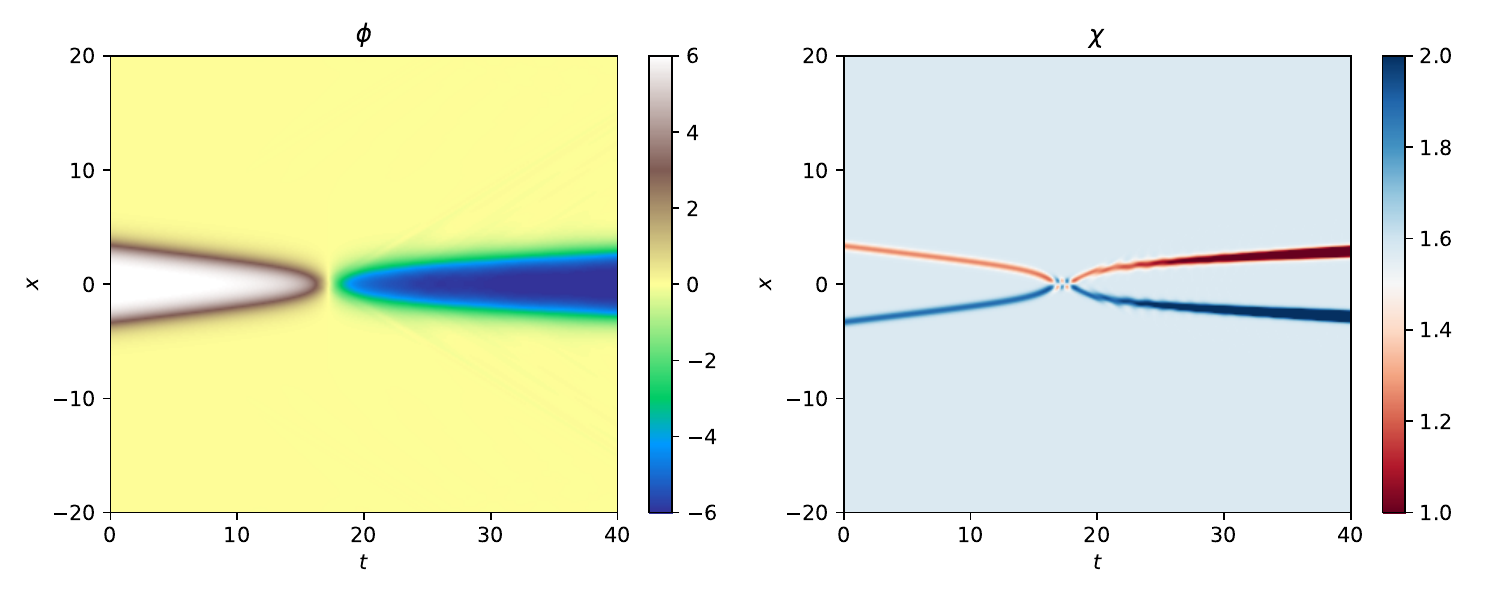}
	 \caption{$K_{0,1}\bar K_{0,-1}$ scattering. Fields (left) $\phi(x,t)$ and (right) $\chi(x,t)$ with $C=0.5$ for a) $v=0.10$, with $C=3.0$ for b) $v=0.18$ and c) $v=0.20$.}
    \label{fig_scat3}
\end{figure}

\begin{figure*}[!ht]
\begin{center}
  \centering
    \subfigure[ ]{\includegraphics[{angle=0,height=6cm,width=17cm}]{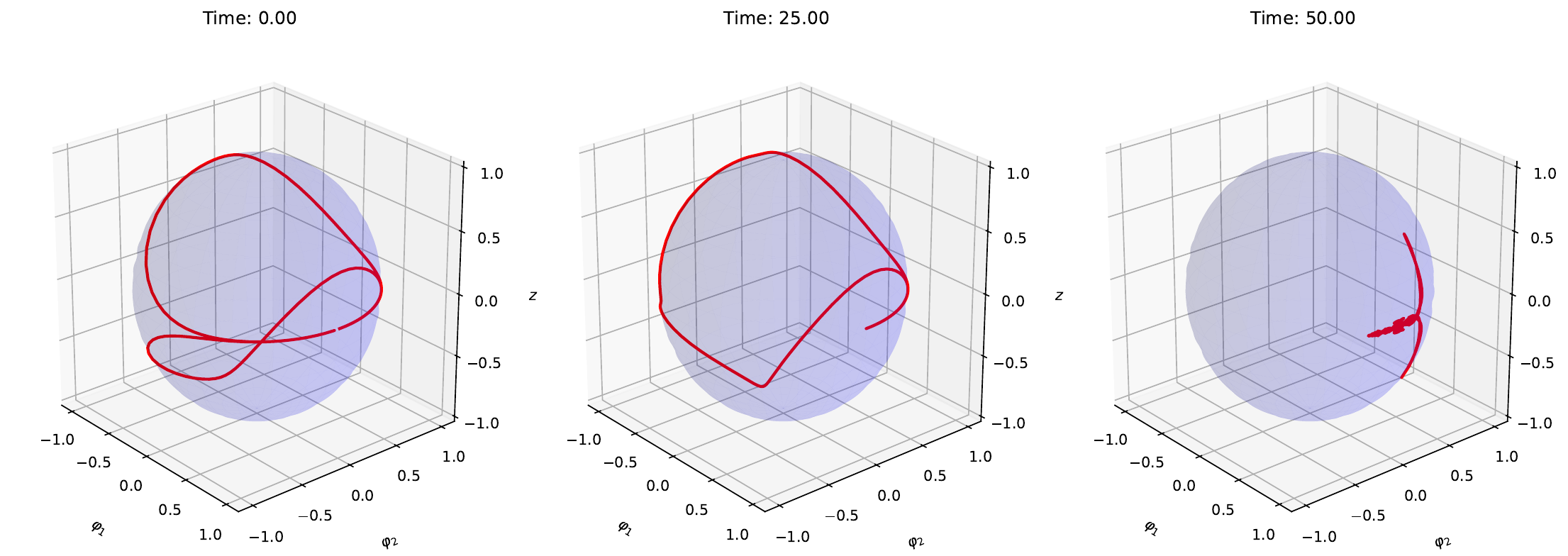}\label{sp_v01_c05}}
    \subfigure[ ]{\includegraphics[{angle=0,height=6cm,width=17cm}]{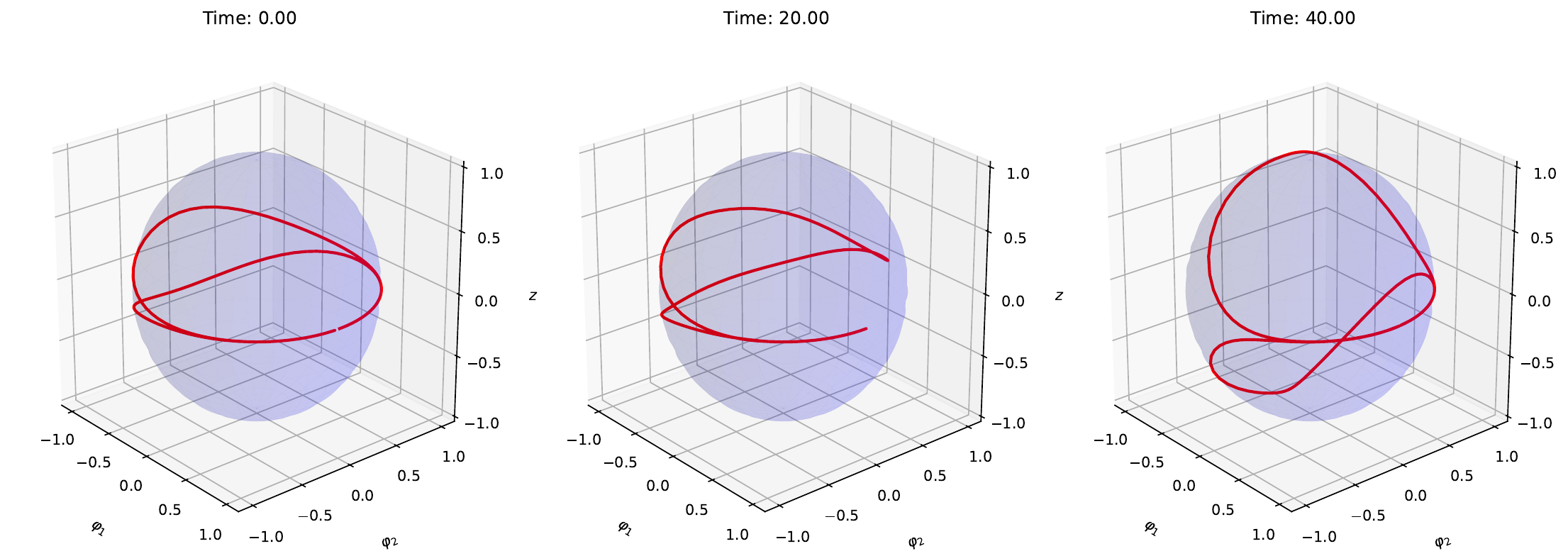}\label{sp_v02_c3}}
        \caption{$K_{0,1}\bar K_{0,-1}$ scattering. Scattering in internal space for a) $C=0.5$ with $v=0.10$ and b) $C=3.0$ with $v=0.20$. Videos from the scattering can be seen, respectively, in the supplementary material video3.mp4 and video4.mp4.} 
	\label{fig_sphere2}
\end{center}
\end{figure*}

In the Fig.~\ref{fig_scat3}a-c we present the results for some values of $C$ and different initial velocities. A new behavior is seen in Fig.~\ref{fig_scat3}a for $C=0.5$, not shown in the former section for $K_{0,1}\bar K_{0,1}$. In the figure one can see that the kink-antikink pair is followed after the collision by radiation jets and two low frequency, in phase oscillating pulses around the vacuum. For the antilump-lump collision, we see  the production of a pair of oscillating pulses with much higher frequencies and opposite phases.  

The increasing of $C$ favors the change of topological sector during the collision process. This is shown in the Figs. ~\ref{fig_scat3}b and ~\ref{fig_scat3}c for $C=3$ and two different initial velocities. In both situations, we see a change in the topological sector with few emitted radiation for the $\phi$ field. However, there are significant differences for the $\chi$ field. For $v=0.18$ (Figs.~\ref{fig_scat3}b) we see a changing in the $\chi$ field from antilump-lump to lump-antilump. However, when the initial velocity increases to $v=0.20$ (Fig.~\ref{fig_scat3}c), the $\chi$ field 
maintain the antilump-lump profile after interaction.

In the Figs. \ref{fig_sphere2}a and \ref{fig_sphere2}b we represent the scattering process in the internal field space. The parameters are, respectively, the same as the Figs. \ref{fig_scat3}a and  \ref{fig_scat3}c. First of all note that the initial configuration corresponds to a set of two loops: one in the north hemisphere, corresponding to $K_{0,1}$ and the other in the south hemisphere,  corresponding to $\bar K_{0,-1}$. These two loops have the point $(1,0,0)$ in common, corresponding to the vacuum for $x\to\pm\infty$. The Fig. \ref{fig_sphere2}a shows that the quasi-annihilation of the $K_{0,1}\bar K_{0,-1}$ pair leads to the collapse of the loops almost to the point corresponding to the vacuum. However, as we saw in the Fig. \ref{fig_scat3}a, each one of the $\phi$ and $\chi$ fields survive as two sets of oscillations with different frequencies, with those from the $\chi$ field around $\chi=\pi/2$ with the higher ones. 
 Also note that, for all time after the scattering, the $\chi$ field in each oscillation has opposed phase with respect to the other oscillation, whereas for the $\phi$ field the two oscillations are in phase.  In the internal space this corresponds to a patter after scattering of two crossed lines around the vacuum (see the third diagram of the Fig. \ref{fig_sphere1}a), one with lower frequency along the equator, and other with higher frequency along the meridians.  
 The Fig. \ref{fig_sphere2}b shows that the changing of topological sector from $(\phi,\chi)=(0,\pi/2)\to(2\pi,\pi/2)$ to $(0,\pi/2)\to(-2\pi,\pi/2)$ (see the Fig. \ref{fig_scat3}c)
does not alter the initial pattern of two loops in separated hemispheres. The final configuration shows that the loops separate further since, according to the Fig. \ref{fig_scat3}c, the final value of $\chi$ field is reduced (for the antilump) and increased (for the lump). 

\section{Stability Analysis}

Stability analysis lead to some interesting hints about soliton-antisoliton scattering. For general $C$, the equations for linear perturbations for the fields $\phi$ and $\chi$ are coupled, as shown in the discussion following the Eq. (\ref{eq_pert}). Luckily, two special cases $C=0$ and $C\to\infty$ are physically interesting, and also allow for a decoupling of such equations. 

Firstly lets us consider $C\ll 1$, corresponding to 
$\chi\sim 0$. For this case  We reported above numerical instabilities. Indeed, from the Fig. \ref{fig_small}a  one can see that in the limit $C\to 0$ one has $\chi=0$ for all finite $x$. Then, Eq. (\ref{eq_pert}) from linear perturbations gives
\be
-\eta''_2+(-V^2_{\phi\chi}/V_{\phi\phi}+V_{\chi\chi}+4\sin^2(\phi/2))\eta_2=\omega^2\eta_2.
\ee
This is a Schrodinger-like equation with a potential unbounded from bellow, meaning instability. This is not in contradiction from the soliton as a stable energy minimized solution. Indeed, the numerical instability occurs when the scalar field $\chi$ goes too nearly to $\chi=0$, an unstable point. 

\begin{figure}
    \includegraphics[{angle=0,width=8cm,height=5cm}]{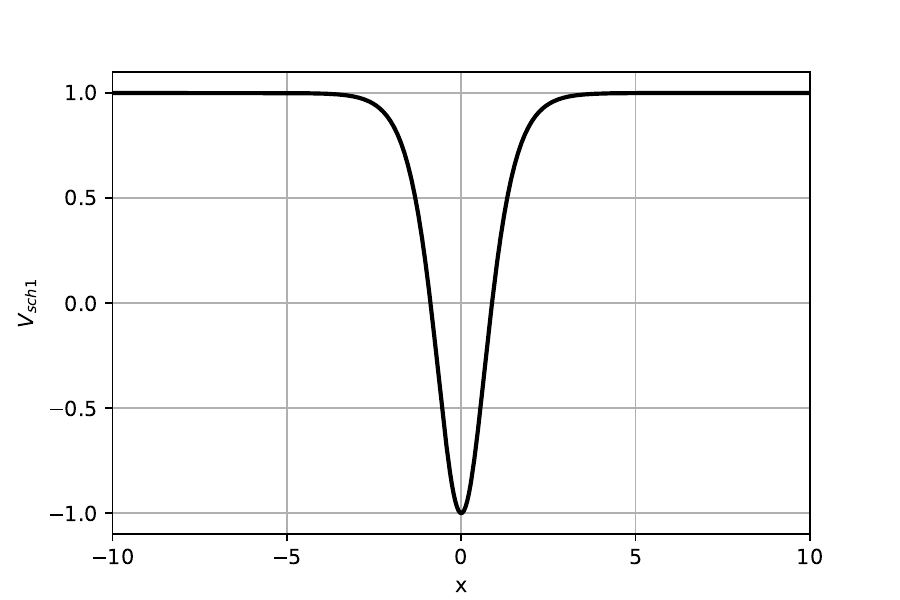}
    \includegraphics[{angle=0,width=8cm,height=5cm}]{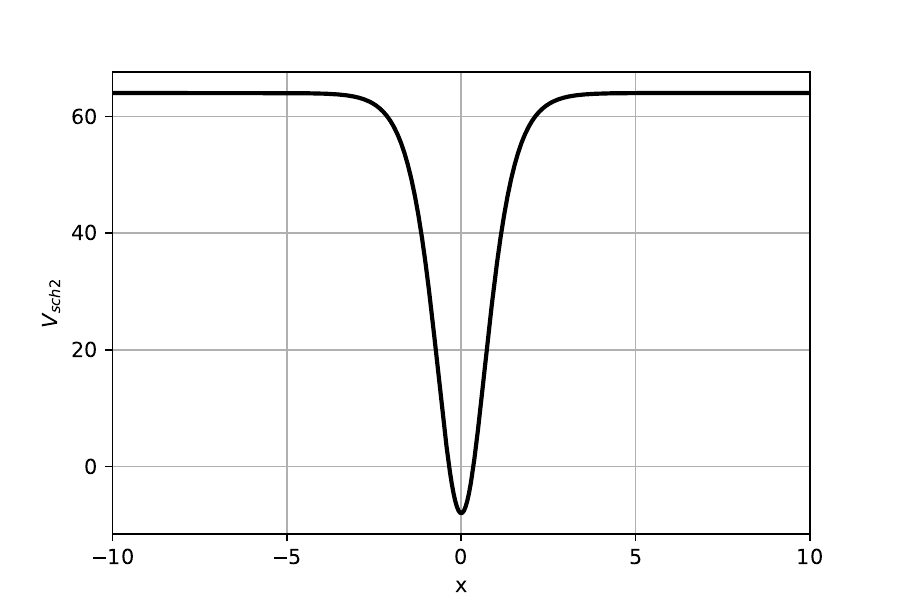}
    \caption{a) Schrodinger-like potentials for the fields $\phi$ (left) and $\chi$ (right) for $C\to\infty$.}
    \label{fig_Vsch_inf}
\end{figure}

On the other hand, the limit $C\to\infty$ results in a stable defect. This can be proved since in this case one has $\chi\to \pi/2$. In this case, the Eq. (\ref{eq_pert}) gives the following perturbation equations:
\begin{eqnarray}
-\eta''_1+V_{\phi\phi}\eta_1&=&\omega^2\eta_1 \label{cinf-a}\\
-\eta''_2+(-\phi'^2+V_{\chi\chi})\eta_2&=&\omega^2\eta_2. \label{cinf-b}
\end{eqnarray}
The Schrodinger potential for the fields $\phi$ and $\chi$ are given, respectively, by $V_{sch1}=V_{\phi\phi}=\cos(\phi(x))$ and $V_{sch2}=-\phi'^2+V_{\chi\chi}=36V_{schr1}+56/2$. The potentials are  presented in the Fig. \ref{fig_Vsch_inf}. Note that both potentials are bounded from bellow, and that there is a linear relation between them. To guarantee stability one must prove that $\omega^2<0$ is forbidden.
Firstly, note from the Eq. (\ref{cinf-a}) that the linear potential for the $\phi$ field is given by $V_{\phi\phi}$. This is exactly the potential of perturbations for the $(1,1)$ integrable sine-Gordon model. This potential has only one mode corresponding to $\omega^2=0$, and no vibrational modes. This is the zero-mode, responsible for the translational invariance. Secondly, from the relation between $V_{sch1}$ and $V_{sch2}$ we see that there is no unstable mode also for the Eq. \ref{cinf-b}, completing the proof.

\section{Conclusion}

In this work we have considered an $(1,1)$ action which is an extension of the nonlinear $O(3)$-sigma model. Our model is characterized by two fields with both geometric and direct coupling with a potential $V(\phi,\chi)$, where the fields are spherical coordinates of the fields $\varphi^i$, with $i=1,2,3$ satisfying the constraint $\vec\varphi\cdot\vec\varphi=1$, where $\vec\varphi=(\varphi^1, \varphi^2, \varphi^3)$, meaning that the internal space of field configuration is that of a $S^2$ surface. In the Ref. \cite{alon2} it was considered the simplest potential $V(\varphi^1,\varphi^2,\varphi^2)$ able to giving mass to the fields, and leading to a massive nonlinear sigma model with several solutions, including two types of topological kinks coming from the twofold embedding of the sine-Gordon model. Here we also considered a massive sigma model, but instead of starting with the potential, as done in \cite{alon2}, we investigated the class of potentials $V(\phi,\chi)$ in the internal curved space $S^2$ that give BPS solution for the $\phi$ field as of a sine-Gordon kink. 

We studied the simplest solution, where  the potential has an infinite number of minima, but only two in the domain $0\leq \phi \leq 2\pi$ and $0\leq \chi \leq \pi$ of $S^2$, meaning one topological sector. For this sector, we can attain two soliton and two antisoliton solutions, where the defect shows a smooth transition between the two minima. 
We found that the $\chi$ field has a lump profile that depends on a constant $C$. The energy density has the profile of a central peak for large $|C|$. On the other hand, for small values of $|C|$, it shows a splitting that grows for even smaller values of $|C|$. This signals that the defect acquires an internal structure for small values of $|C|$. In the field space of the fields $(\varphi_1,\varphi_2,\varphi_3)$ we showed that the defects are closed loops in $S^2$. The loops are in the north hemisphere for $C>0$ and in the south hemisphere for $C<0$. The higher is $|C|$, the closer is the loop to the equator. Reducing the value of $|C|$ means loops that interpolate between the equator an lower (for $C>0$) or higher (for $C<0$) latitudes. We investigated two possibilities, for the defect characterized by the fields $(\phi,\chi)$ in the form of kink-antilump: i) scattering with an antikink-antilump and ii)  i) scattering with an antikink-lump. Depending on $C$ and the initial velocity $v$, we identified the main results for scattering: 
one-bounce scattering for $\phi$, or strong emission of radiation for $\phi$, followed by  i) annihilation of $\chi$; ii) same pattern antilump-antilump or lump-antilump for $\chi$; iii) inversion antilump-antilump to lump-lump for $\chi$; iv) inversion antilump-lump to lump-antilump for $\chi$.
Other findings are: v) annihilation of the pair soliton-antisoliton with the emission of scalar radiation; vi) emission of pairs of oscillations around the vacuum for $\phi$ and $\chi$. We observed that the increasing of $|C|$ leads to the reduction of radiation emission.

We studied some aspects form the dynamics in the internal space, were the initial configuration of the soliton-antisoliton pair corresponds to two loops that can be both in the north hemisphere (for $K_{0,1}\bar K_{0,1}$) or in separated hemispheres (for $K_{0,1}\bar K_{0,-1}$). After scattering, the loops can develop interconnections in a complex pattern or be transformed in open strings. However, we did not observe the production of separated loops or strings. Indeed, whatever is the scenario, the  point $(\varphi^1,\varphi^2,\varphi^3)=(1,0,0)$ (the vacuum) always belongs to the string or loop. 

The BPS solutions are, by construction, solutions which minimize energy. Then they are stable solutions, but difficult to analyze due to the coupling between the linear perturbation equations. We showed that for $|C|\to\infty$ and $C=0$ the linear perturbation equations decouple and are reduced to a Sturm-Liouville problem, whose solutions have a simple interpretation.
Also, despite static solutions for general $C$ being energetically stable, the study of scattering of soliton-antisoliton for small $C$ suffer from numerical instability. There is no contradiction here, since this is shown to be the result of the instability for $\chi=0$. Interestingly, the limit $|C|\to\infty$ corresponds to the sine-Gordon solution for $\phi$, with $\chi=\pi/2$. In this limit it is expected that the scattering between the defects to be elastic, as occurs with the integrable $(1,1)$ sine-Gordon model.    In this way, our numerical investigation shows how a nonintegrable model constructed with two fields behaves due to scattering as one approaches the conditions for integrability.

\section{Supplementary Material}

\begin{itemize}
\item \href{link here}{\texttt{video1.mp4}} - $K_{0,1}\bar{K}_{0,1}$ scattering: solution in internal space for $v=0.20$ with $C=5.0$.
\item \href{link here}{\texttt{video2.mp4}} - $K_{0,1}\bar{K}_{0,1}$ scattering: solution in internal space for $v=0.40$ with $C=5.0$.
\item \href{link here}{\texttt{video3.mp4}} - $K_{0,1}\bar{K}_{0,-1}$ scattering: solution in internal space for $v=0.10$ with $C=0.5$.
\item \href{link here}{\texttt{video4.mp4}} - $K_{0,1}\bar{K}_{0,1}$ scattering: solution in internal space for $v=0.20$ with $C=3.0$.
\end{itemize}


\section*{Acknowledgements}

A.R. Gomes thanks R. Casana for discussions, FAPEMA - Funda\c c\~ao de Amparo \`a Pesquisa e ao Desenvolvimento 
do Maranh\~ao through Grants Universal
and 01441/18 and CNPq (brazilian agency) through Grants 313014. This study was financed in part by
the Coordena\c c\~ao de Aperfei\c coamento de Pessoal de
N\'ivel Superior - Brasil (CAPES) - Finance Code 001 for financial support.


\bibliographystyle{elsarticle-harv} 




\end{document}